\documentclass[twocolumn,english,aps,prl,twocolumn,showpacs,superscriptaddress,groupaddress,floatfix,graphics,graphicx]{revtex4-1}
\usepackage[T1]{fontenc}
\usepackage[latin9]{inputenc}
\usepackage{color}
\usepackage{babel}
\usepackage{textcomp}
\usepackage{amsmath}
\usepackage{amssymb}
\usepackage{graphicx}
\usepackage{esint}
\usepackage[unicode=true,pdfusetitle,
 bookmarks=true,bookmarksnumbered=false,bookmarksopen=false,
 breaklinks=false,pdfborder={0 0 1},backref=false,colorlinks=false]
 {hyperref}
\hypersetup{
 colorlinks,linkcolor=blue,citecolor=blue,urlcolor=blue}

\makeatletter

\newcommand*\LyXThinSpace{\,\hspace{0pt}}

\usepackage{babel}

\date{\today}

\makeatother

\begin{document}
\title{Spin Hall and inverse spin galvanic effects in graphene with strong
interfacial spin-orbit coupling: a quasi-classical Green's function
approach}
\author{Carmen Monaco }
\affiliation{Dipartimento di Matematica e Fisica, Università  Roma Tre, via della
Vasca Navale 84, 00146 Rome Italy}
\author{Aires Ferreira}
\affiliation{Department of Physics and York Centre for Quantum Technologies, University
of York, York YO10 5DD, United Kingdom}
\author{Roberto Raimondi}
\affiliation{Dipartimento di Matematica e Fisica, Università  Roma Tre, via della
Vasca Navale 84, 00146 Rome Italy}
\begin{abstract}
\textcolor{black}{van der Waals heterostructures assembled from atomically
thin crystals are ideal model systems to study spin-orbital coupled
transport because they exhibit a strong interplay between spin, lattice
and valley degrees of freedom that can be manipulated by strain, electric
bias and proximity effects. The recently predicted spin-helical regime
in graphene on transition metal dichalcogenides, in which spin and
pseudospin degrees of freedom are locked together {[}M. Offidani}\textcolor{black}{\emph{
et al}}\textcolor{black}{. Phys. Rev. Lett. 119, 196801 (2017){]},
suggests their potential application in spintronics. Here, by deriving
an Eilenberger equation for the quasiclassical Green's function of
two-dimensional Dirac fermions in the presence of} spin-orbit coupling\textcolor{black}{{}
(SOC) and scalar disorder, we obtain analytical expressions for the
dc spin galvanic susceptibility and spin Hall conductivity in the
spin-helical regime. Our results disclose a sign change in the spin
Hall angle (SHA) when the Fermi energy relative to the Dirac point
matches the Bychkov-Rashba energy scale, irrespective of the magnitude
of the spin-valley interaction imprinted on the graphene layer. The
behavior of the SHA is connected to a reversal of the total internal
angular momentum of Bloch electrons that reflects the spin-pseudospin
entanglement induced by SOC. We also show that the charge-spin conversion
reaches a maximum when the Fermi level lies at the edge of the spin-minority
band in agreement with previous findings. Both features are fingerprints
of spin-helical Dirac fermions and suggest a direct way to estimate
the strength of proximity-induced SOC from transport data. The relevance
of these findings for interpreting recent spin-charge conversion measurements
in nonlocal spin-valve geometry is also discussed.}
\end{abstract}
\maketitle

\section{Introduction}

van der Waals heterostructures \citep{Geim_2013} have become one
of the most promising spintronic platforms, where both fundamental
and applied aspects of spin transport can be addressed with exquisite
electrical control in the atomically thin limit \citep{Han_2014,Avsar_2020,Cavill_2020}. 

Soon after graphene became well established as a high-performance
spin channel supporting spin transport over long distances at room
temperature \citep{Tombros_2007,Guimaraes_2012,Kamalakar_2015,Drogeler_16,Gebeyeue_2019},
the primary focus has shifted towards the study of emergent spin-charge
coupling effects in van der Waals heterostructures. An intriguing
possibility consists of exploiting relativistic SOC phenomena to generate
and manipulate spin-polarization flow in atomically thin planes. One
of the first proposed schemes made use of proximity-induced SOC in
graphene flakes with a dilute coverage of heavy adatoms \citep{Ferreira_2014,Balakrishnan_2014,Tuan_2016},
which are believed to be efficient extrinsic sources of spin Hall
currents and nonequilibrium spin polarization \citep{Pachoud_2014,Huang_2016}.
An alternative approach, which recently meets a lot of attention,
consists of enhancing the SOC by placing the graphene sheet on top
of a layered semiconductor \citep{Avsar_2014,Wang_2015,Gmitra_2015,Offidani_2017,Island_19,Wang_21,Lin_21}.
\textcolor{black}{It is now understood that the breaking of inversion
symmetry in van der Waals heterostructures results in dramatically
enhanced intrinsic- and Bychkov-Rashba (BR)-type SOC \citep{Bychkov_1984,Rashba_2009,Rashba_2012},
endowing spin-split Dirac cones with a robust skyrmion-like spin texture
in $\mathbf{k}$ space \citep{Offidani_2018c}.} The interface-induced
SOC, whose precise spatial profile reflects the interlayer atomic
registry and disorder landscape, can be further manipulated by applying
strain and electric fields \citep{random_SOC_Hernando_09,Gmitra_2016,Santos_2018,Cysne_2018,Xu_18,Leutenantsmeyer_18}.\textcolor{blue}{{}
}Low-temperature magnetotransport data for graphene on transition
metal dichalcogenides (TMDs) is consistent with interface-induced
SOC in the range 1--10 meV \citep{WAL_Volkl_2017,WAL_Wakamura_2018,WAL_Wang_2016,WAL_Yang_2017,WAL_Zihlmann_18},
up to two orders of magnitude higher than graphene's weak intrinsic
SOC ($\lambda_{I}\simeq42.2$ $\mu$eV \citep{Sichau_19}), in good
agreement with theoretical predictions based on density functional
theory and semi-empirical methods \citep{Wang_2015,Gmitra_2016,Cysne_2018}.
Concurrently, room-temperature Hanle-type spin precession measurements
revealed another fingerprint of proximitized graphene, that is, a
giant ratio of out-of-plane to in-plane spin lifetimes ($\tau_{\perp}/\tau_{\parallel}\approx10$
\citep{Benitez_2017,Ghiasi_2017,Zihlmann_2018}) driven by the competition
of symmetry distinct spin-orbit interactions and intervalley scattering
\citep{Cummings_2017,Offidani_2018}. Meanwhile, quantum Hall effect
measurements performed on ultra-clean bilayer graphene/few-layer WSe$_{2}$
have shown that interfacial SOC can be made as large as 15 meV by
removing contaminants from the device areas \citep{Wang_21}. 

\textcolor{black}{A high interfacial SOC with magnitude comparable
to the quasiparticle lifetime broadening is a desirable feature because
it allows efficient spin-charge conversion via spin-helical Dirac
fermions \citep{Offidani_2017} as demonstrated in a series of elegant
spin-valve experiments on graphene/TMD bilayers \citep{Safeer_2019,Ghiasi_2019,Benitez_2020,Khokhriakov_2020,Lin_2020}.}
The delicate interplay between intrinsic and extrinsic (impurity-driven)
spin-orbit-coupled transport mechanisms in graphene-based heterostructures
has been recently studied by means of a linear response formalism,
supported by conservation laws \citep{Milletari_2016a,Milletari_2016b,Milletari_2017,Offidani_2018b}.
Unlike BR-coupled two-dimensional (2D) electron gases \citep{Inoue_04,Mishchenko_04,Dimitrova_05},
the spin Hall conductivity in an infinite system with nonmagnetic
defects was found to be finite due to the generally noncoplanar nature
of the equilibrium spin texture at the Fermi energy \citep{Milletari_2017}.
The critical role played by impurity scattering in the context of
SOC-driven spin-charge conversion has also been investigated by means
of the non-equilibrium Green's function technique, which is particularly
suited to derive coupled spin-charge drift-diffusion equations \citep{Raimondi_2006,Gorini_2010,Raimondi_2012}.
In particular, the Keldysh technique in the so-called quasiclassical
approximation, pioneered by Eilenberger \citep{Eilenberger_1968, Rammer_1986}
for dirty type II superconductors, has been applied to describe the
locked spin-charge dynamics of topological insulators (TI) \citep{Schwab_2011}. 

The aim of the present paper is to extend the quasi-classical approach
developed in Refs. \citep{Raimondi_2006,Gorini_2010,Raimondi_2012}
to a system of dirty 2D Dirac fermions subject to strong proximity-induced
SOC. Our focus will be on the low-density regime highlighted in Ref.
\citep{Offidani_2017}, in which the Fermi energy crosses a single
spin-split band, and thus the 2D Dirac fermions acquire a well-defined
spin helicity akin to surface states of a three-dimensional topological
insulator. \textcolor{black}{Our work is organized as follows: In
Section II we introduce the effective Hamiltonian of graphene/TMD
heterostructures and discuss how proximity-induced SOC modifies the
electronic structure at low energies. In Section III, we present the
Keldysh technique and in Section IV we derive the Eilenberger equation
for the quasiclassical Green's function in the spin-helical regime.
In particular we discuss the $\mathrm{T}-$matrix expansion for the
disorder potential and derive the general expression for the collision
integral. In Section V, by confining to the Born approximation for
the self-energy, i.e. the second order in the $\mathrm{T}-$matrix
expansion, we solve the Eilenberger equation and find the expressions
for the longitudinal electrical conductivity and the spin galvanic
susceptibility, while there is no spin Hall effect due to the absence,
at this order, of skew scattering. The latter is explicitly considered
in Section VI by carrying out the T-matrix expansion up to the third
order in the scattering potential. (Some technical details are relegated
to the Appendix for clarity of exposition.) The explicit solution
of the Eilenberger equation provides the expressions of the electrical
conductivity, spin galvanic susceptibility and spin Hall conductivity
in terms of the dimensionless disorder coupling strength and of the
energy eigenstate at the Fermi level. Recent spin precession measurements
of inverse spin Hall and spin galvanic effects in a graphene/WS$_{2}$
heterostructure \citep{Benitez_2020} are put into context. Finally
Section VII presents our conclusions. }

\section{The model}

\begin{figure}
\includegraphics[width=1\columnwidth]{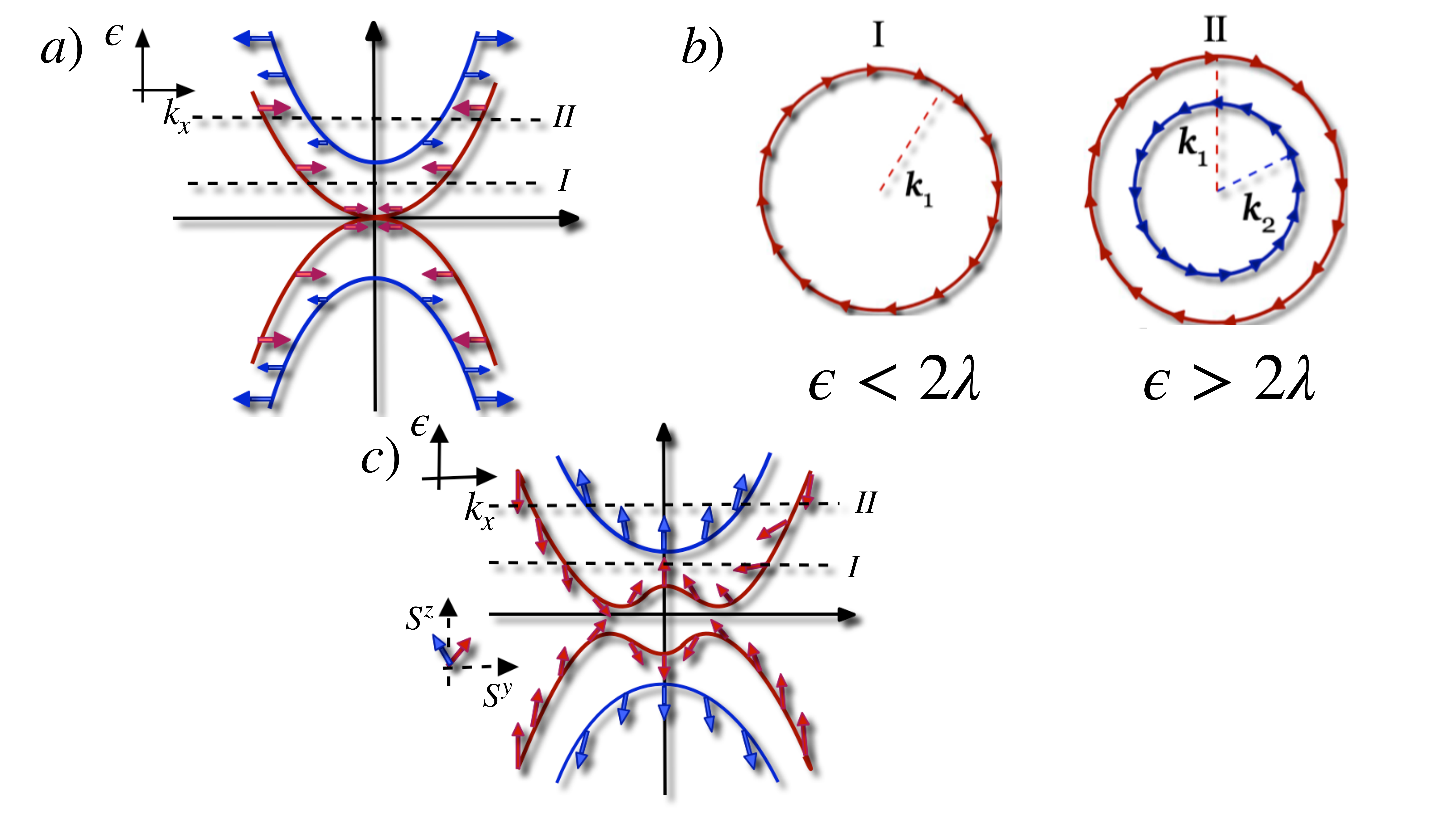} \caption{(a-b) Energy dispersion of the Dirac-Rashba model and corresponding
equilibrium spin texture in $C_{\textrm{6v}}$-symmetric heterostructures
(where $\lambda_{\textrm{sv}}=0$). The spin-splitting of the Dirac
bands leads to a spin gap of width $2\lambda$, which separates a
2D electron gas-like region of BR-split bands with counter-rotating
spin textures (regime II: $\epsilon>2\lambda$) from a spin-helical
regime (regime I: $\epsilon<2\lambda$). (c) Same as (a) for graphene
heterostructures with $C_{3v}$ point group symmetry. The competition
of BR and spin-valley SOC gives rise to a ``Mexican hat-shaped'' dispersion
with an electron/hole pocket at very low energies. Energies are defined
with respect to the Dirac point. For visualisation purposes, the bands
are plotted along $k_{x}$ with $k_{y}=0$ (i.e. spins lie only in
the $yz$ plane). }
\label{fig1} 
\end{figure}

The low-energy excitations in graphene/TMD bilayers are governed by
the following generalized Dirac-Rashba model (we use natural units
$\hbar=1$ throughout)

\begin{equation}
H_{\xi}=\intop d\mathbf{x}\,\Psi_{\xi}^{\dagger}\left[-\imath\mathit{v}\,\boldsymbol{\sigma}\cdot\mathbf{\boldsymbol{\nabla}}+V_{\mathit{\textrm{SO}}}^{\xi}+U\left(\mathbf{x}\right)\right]\Psi_{\xi}\,,\label{eq:ham}
\end{equation}
where $\xi=\pm1$ signs refer to inequivalent valleys $K(+)$ and
\textbf{$K^{\prime}(-)$}, $(\Psi_{\xi},\Psi_{\xi}^{\dagger})\equiv(\Psi_{\xi}(\mathbf{x}),\Psi_{\xi}^{\dagger}(\mathbf{x}))$
are 4-component spinor fields defined on the internal spaces of pseudospin
and spin and $v$ is the velocity of massless Dirac fermions (here,
$\boldsymbol{\sigma}$ is the vector of Pauli matrices acting on the
pseudospin space). The term $V_{\mathit{\textrm{SO}}}^{\xi}$ describes
the spatially uniform proximity-induced SOC on the graphene layer
and comprises several contributions that reflect the $C_{3v}$ point
group symmetry of the heterostructure \citep{Wang_2015,Fabian_2017,Cysne_2018}.
For instance, the breaking of mirror reflection symmetry about the
graphene plane allows $z\rightarrow-z$ asymmetric SOC. This so-called
BR effect \citep{Bychkov_1984} is generically present in graphene
on a substrate and is responsible for the entanglement between pseudospin
and spin degrees of freedom \citep{Rashba_2009}, with clear signatures
in the spin dynamics \citep{Tuan_2014} and current-induced spin-orbit
torque \citep{Sousa_2020}. Moreover, the spin-orbit interactions
imprinted on the graphene layer acquire sublattice-resolved terms
inherited from the noncentrosymmetric TMD layer, namely the spin-valley
coupling, as mentioned in the Introduction. All together, in the longwavelenght
limit, $V_{\mathit{\textrm{SO}}}^{\xi}$ generally contains three
SOC terms compatible with the $C_{3v}$ point group i.e.,

\begin{equation}
V_{\mathit{\textrm{SO}}}^{\xi}=V_{\textrm{KM}}+V_{\textrm{BR}}+V_{\textrm{sv}}^{\xi},
\end{equation}

\noindent where $V_{\textrm{KM}}\propto\sigma_{z}s_{z}$ is the intrinsic-type
Kane-Mele SOC \citep{KaneMele_2005,KaneMele_2005b}, $V_{\textrm{BR}}\propto(\sigma_{x}s_{y}-\sigma_{y}s_{x})$
describes the BR effect, \citep{Rashba_2009} $V_{\textrm{sv}}^{\xi}\propto\xi s_{z}$
captures the spin-valley effect which acts on charge carriers as a
valley-Zeeman coupling \citep{Srivastava_2015}, and $\mathbf{s}$
is the vector of Pauli matrices acting on the spin space. In practice,
the weak $z\rightarrow-z$ symmetric Kane-Mele SOC can be neglected
in comparison to the other effects (see, for example, Ref.~\citep{Gmitra_2016})
and hence in this work, we approximate the spin-orbit interaction
as follows

\begin{equation}
V_{\mathit{\textrm{SO}}}^{\xi}\simeq\lambda\left(\boldsymbol{\sigma}\times\mathbf{s}\right)_{z}+\xi\lambda_{\textrm{sv}}s_{z}.\label{eq:SOC}
\end{equation}
Beyond SOC, the proximity coupling to a TMD also imprints a sublattice-staggered
potential $H_{\Delta}^{\xi}=\Delta\xi\sigma_{z}$ on the graphene
sheet. The staggered on-site energy is believed to be substantially
smaller than the SOC energy scales in Eq.\,(\ref{eq:SOC}) (under
0.1 meV according to a recent study \citep{Pezo_2020}) and because
it plays a very limited role in both the spin dynamics \citep{Offidani_2018}
and coupled spin-charge transport effects \citep{Offidani_2017},
we neglect it in the following discussion. We note that more SOC terms
are allowed if additional crystal symmetries are broken further reducing
the point group of the heterointerface. These include unconventional
in-plane-polarized spin-valley and Kane-Mele-type SOC that are symmetry-allowed
in graphene coupled to low-symmetry TMDs \citep{Cysne_2018}, the
implications of which are beyond the scope of the present work. Finally,
the last term in Eq.(\ref{eq:ham}) is a random potential produced
by scalar impurities, which will be responsible for the extrinsic
generation of nonequilibrium spin polarization and spin Hall currents
to be discussed in Secs. \ref{sec:IV}-\ref{sec:VI}.

The energy-momentum dispersion relation of the low-energy Dirac bands
reads as

\begin{equation}
\epsilon_{ln}\left(\mathbf{k}\right)=l\sqrt{v^{2}k^{2}+[M_{n}(k)]^{2}},\label{eq:eigen}
\end{equation}

\noindent where $M_{n}(k)=[2\lambda^{2}+\lambda_{\textrm{sv}}^{2}+2n\sqrt{\lambda^{4}+(\lambda^{2}+\lambda_{\textrm{sv}}^{2})v^{2}k^{2}}]^{1/2}$
is the spin gap induced by SOC, $n,l=\pm1$ are the band indices and
$k=|\mathbf{k}|$ (with $\mathbf{k}$ measured from a $K$ point).
Equation (\ref{eq:eigen}) makes manifest the underlying particle-hole
symmetry of the Hamiltonian, which results in one or two Dirac bands
at the Fermi energy $\varepsilon$ depending on the gate-tunable charge
carrier density (see Fig. \ref{fig1}). To ease the notation, we shall
assume $\epsilon,\lambda_{\textrm{sv}},\lambda>0$ in what follows.
Inverting Eq.(\ref{eq:eigen}) and evaluating the energy-momentum
dispersion relation at the Fermi energy $\epsilon_{ln}=\varepsilon$,
one obtains the Fermi momenta, $k_{ln}=k(\epsilon_{ln}=\varepsilon)$,
as follows

\begin{equation}
k_{ln}=v^{-1}\sqrt{\lambda_{\textrm{sv}}^{2}+\varepsilon^{2}\pm2\sqrt{\lambda^{2}\varepsilon^{2}+\lambda_{\textrm{sv}}^{2}\varepsilon^{2}-\lambda^{2}\lambda_{\textrm{sv}}^{2}}}.\label{eq:kln}
\end{equation}
The $\pm$ sign depends on the energy range within which lies the
Fermi energy $\varepsilon$ (Fig. \ref{fig1}). The eigenvectors,
$\psi_{ln}(\mathbf{x})=e^{i\mathbf{k}\cdot\mathbf{x}}\Phi_{ln}(\mathbf{k})$,
have the following spinorial structure

\begin{equation}
\Phi(k_{ln},\theta)=\left(\begin{array}{c}
e^{-i\theta}\\
i\alpha\\
\beta\\
i\gamma e^{i\theta}
\end{array}\right)\label{eq:generic_eigenvector}
\end{equation}

\noindent with $\theta$ the momentum angle with respect to an arbitrary
axis in the graphene plane and $\alpha,\beta$ and $\gamma$ given
by 
\begin{eqnarray}
\alpha & = & \frac{(\epsilon_{ln}-\lambda_{\textrm{sv}})^{2}-v^{2}k_{ln}^{2}}{2\lambda vk_{ln}},\label{eq:alpha}\\
\beta & = & \frac{\epsilon_{ln}-\lambda_{\textrm{sv}}}{vk_{ln}},\label{eq:beta}\\
\gamma & = & \frac{(\epsilon_{ln}-\lambda_{\textrm{sv}})^{2}-v^{2}k_{ln}^{2}}{2\lambda(\epsilon_{ln}+\lambda_{\textrm{sv}})}.\label{eq:gamma}
\end{eqnarray}
In this work we specialize to the regime I which should be experimentally
accessible in ultra-clean devices with strong interfacial SOC. The
Fermi energy therefore lies in the interval $\varepsilon_{-}\equiv\lambda_{sv}<\varepsilon<\sqrt{4\lambda^{2}+\lambda_{sv}^{2}}\equiv\varepsilon_{+}$
{[}Fig. \ref{fig1} (c){]}, which we call spin-helical regime, where
the simply-connected Fermi surface topology is akin to ideal topological
surface states \citep{Schwab_2011}. For that reason, we will drop
the band indices and define the Fermi momentum in regime I as $k_{F}\equiv k_{1,-1}$
with $k_{1,-1}=k_{1,-1}(\varepsilon)$ as per Eq.(\ref{eq:kln}).

For the transport calculations below, we will also need the density
of states at the Fermi level, per valley,

\begin{equation}
N_{F}=\frac{\varepsilon(\lambda^{2}+\lambda_{\textrm{sv}}^{2}+\sqrt{\lambda^{2}\varepsilon^{2}+\lambda_{\textrm{sv}}^{2}(\varepsilon^{2}-\lambda^{2})})}{2\pi v^{2}\sqrt{\varepsilon^{2}(\lambda^{2}+\lambda_{\textrm{sv}}^{2})-\lambda^{2}\lambda_{\textrm{sv}}^{2}}}.\label{eq:dos_3}
\end{equation}

\noindent From this expression, one easily recovers pristine graphene's
well-known expression $N_{0}=\varepsilon/(2\pi v^{2})$ by letting
$\lambda_{\textrm{sv}}\rightarrow0$ first and then $\lambda\rightarrow0$.
We will show below that in this regime, where the electronic states
have a well-defined spin helicity, the pseudospin and spin degrees
of freedom are constrained in such a way that it becomes possible
a full description of the coupled spin-charge dynamics in terms of
a single transport equation. 

\section{The Keldysh technique\label{sec:III}}

The Keldysh formalism, which goes back to the pioneering works by
Schwinger \citep{Schwinger_1961} and Keldysh \citep{Keldysh_1965},
is a powerful generalization of the standard perturbative approach
of equilibrium quantum field theory to nonequilibrium problems. Within
the Keldysh technique \citep{Kadanoff_1962,Raimondi_2003,Rammer_2007,Kamenev_2009,Raimondi_2016},
the Green's function has the following matrix structure 
\begin{equation}
\check{G}=\left(\begin{array}{cc}
G^{R} & G^{K}\\
0 & G^{A}
\end{array}\right),\label{eq:Gmatrix_keldish}
\end{equation}
with $G^{R},G^{A}$ and $G^{K}$ the respective retarded, advanced
and Keldysh components. The Green's function acts on the spin, valley
and pseudospin spaces, and thus each block in Eq. (\ref{eq:Gmatrix_keldish})
can be represented as a $8\times8$ matrix. $\check{G}$ satisfies
the left-right subtracted Dyson equation \citep{Rammer_1986}, which
conveniently gets rid of the delta-function singularity at coinciding
space-time points, i.e.

\begin{equation}
\left[\left(\check{G_{0}}\right)^{-1},\check{G}\right]=\left[\check{\Sigma},\check{G}\right],\label{eq:LR_Dyson}
\end{equation}
where the square brackets define the commutator. Here the products
are to be understood with respect to both the Keldysh and internal
spin spaces and to the space-time coordinates (i.e. convolution products).
The kinetic and proximity-induced SOC terms are contained in the bare
Green's function $\check{G_{0}}$, whereas the quasiparticle self-energy
$\check{\Sigma}$ is due to the disorder potential. In the following
the self-energy will be obtained by averaging, term by term, the perturbative
expansion of the left-right subtracted Dyson equation in the potential
$U\left(\mathbf{x}\right)$ with respect to the disorder configuration.
The Green's function $\check{G}$ entering Eq.(\ref{eq:LR_Dyson})
is then the disorder-averaged Green's function. In explicit terms,
we have

\[
\left(\check{G}_{0}\left(X_{1};X_{2}\right)\right)^{-1}=\left(i\partial_{t_{1}}-h\left(\mathbf{x_{1}}\right)\right)\delta\left(t_{1}-t_{2}\right)\delta\left(\mathbf{x_{1}-\mathbf{x_{2}}}\right)\,,
\]
with $X_{1}\equiv\mathbf{\mathbf{(x_{1}},}t_{1})$, $X_{2}\equiv\mathbf{\mathbf{(x_{2}},}t_{2})$
and $h\left(\mathbf{x}\right)$ the Hamiltonian density as derived
from the disorder-free part in Eq. (\ref{eq:ham}). The externally
applied (uniform) electric field $\mathbf{E}$ is incorporated by
means of the standard minimal coupling within the velocity gauge,
$-i\boldsymbol{\nabla}\rightarrow-i\boldsymbol{\nabla}-e\mathbf{E}\,t$
\citep{Rammer_1986,Rammer_2007}, $t$ being the center-of-mass time
defined below and $e>0$ the unit of electric charge.

The aim of this work is to derive a coupled spin-charge transport
equation in the spin-helical regime. Following the standard approach,
we define the center-of-mass and relative space-time coordinates 
\begin{equation}
\mathbf{x}=\frac{1}{2}\left(\mathbf{x_{1}+\mathbf{x_{2}}}\right),\quad t=\frac{1}{2}\left(\mathbf{\mathrm{t}_{\mathrm{1}}+\mathrm{t}_{\mathrm{2}}}\right)\,,\label{eq:cm_coord}
\end{equation}
\begin{equation}
\mathbf{r}=\mathbf{x}_{1}-\mathbf{x}_{2},\qquad\tau=t_{1}-t_{2}\,.\label{eq:rel_coord}
\end{equation}
As customary, we introduce Wigner-mixed coordinates by taking the
Fourier transform with respect to the $\mathbf{r}$ and $\tau$ variables.
The key assumption in the derivation of the transport equation is
that the center-of-mass space-time variable $\mathbf{x},t$ is a slow
variable compared to $\mathbf{r},\tau$. As a result one can perform
a gradient expansion (see Appendix A for details) of Eq.(\ref{eq:LR_Dyson})
obtaining for $\check{G}\left(\mathbf{x},t,\mathbf{k},\omega\right)\equiv\check{G}$
the following equation of motion 
\begin{eqnarray}
\partial_{t}\check{G}+\frac{1}{2}\left\{ \boldsymbol{\sigma\cdot},\left(\nabla-e\mathbf{E\partial_{\omega}}\right)\check{G}\right\} +\qquad\nonumber \\
+i\left[h\left(\mathbf{k}\right),\check{G}\right]-\frac{1}{2}\left\{ e\mathbf{E\cdot},\nabla_{\mathbf{k}}\check{G}\right\} =-i\left[\check{\Sigma},\check{G}\right],\label{eq:gradient_expansion}
\end{eqnarray}
where $h\left(\mathbf{k}\right)$ is the Fourier-transformed bare
Hamiltonian density and the curly brackets denote, as usual, the anticommutator.
In a compact form, Eq. (\ref{eq:gradient_expansion}) provides the
equation of motion for all the components of the Green's function
Eq.(\ref{eq:Gmatrix_keldish}). The Keldysh component in the right
hand side term of Eq.(\ref{eq:gradient_expansion}) is usually named
collision integral and, in the spirit of\textcolor{black}{{} the }semiclassical
Boltzmann transport theory, it can be divided in a $in$- and $out$-term
according to 
\begin{align}
I & =-i\left(\Sigma^{R}G^{K}-G^{K}\Sigma^{A}\right)+\nonumber \\
 & +i\left(G^{R}\Sigma^{K}-\Sigma^{K}G^{A}\right)\equiv I_{out}+I_{in}\,.\label{eq:collision_integral}
\end{align}

The retarded and advanced clean Green's functions are given by

\begin{equation}
G_{0\mathbf{k}}^{R,A}=\sum_{l,n}\frac{P_{ln}\left(\mathbf{k}\right)}{\omega-\epsilon_{ln}\left(\mathbf{k}\right)\pm i0^{+}},\label{eq:GR}
\end{equation}

\noindent where $P_{ln}\left(\mathbf{k}\right)=|\Phi(k_{ln},\theta)\rangle\langle\Phi(k_{ln},\theta)|$
is the projector onto the eigenstate with indices $ln$. In the spin-helical
regime only the projector $P_{1-1}\left(\mathbf{k}\right)\equiv P\left(\mathbf{k}\right)$
is relevant and we omit the band indices hereafter for simplicity.

\begin{figure}
\includegraphics[scale=0.28]{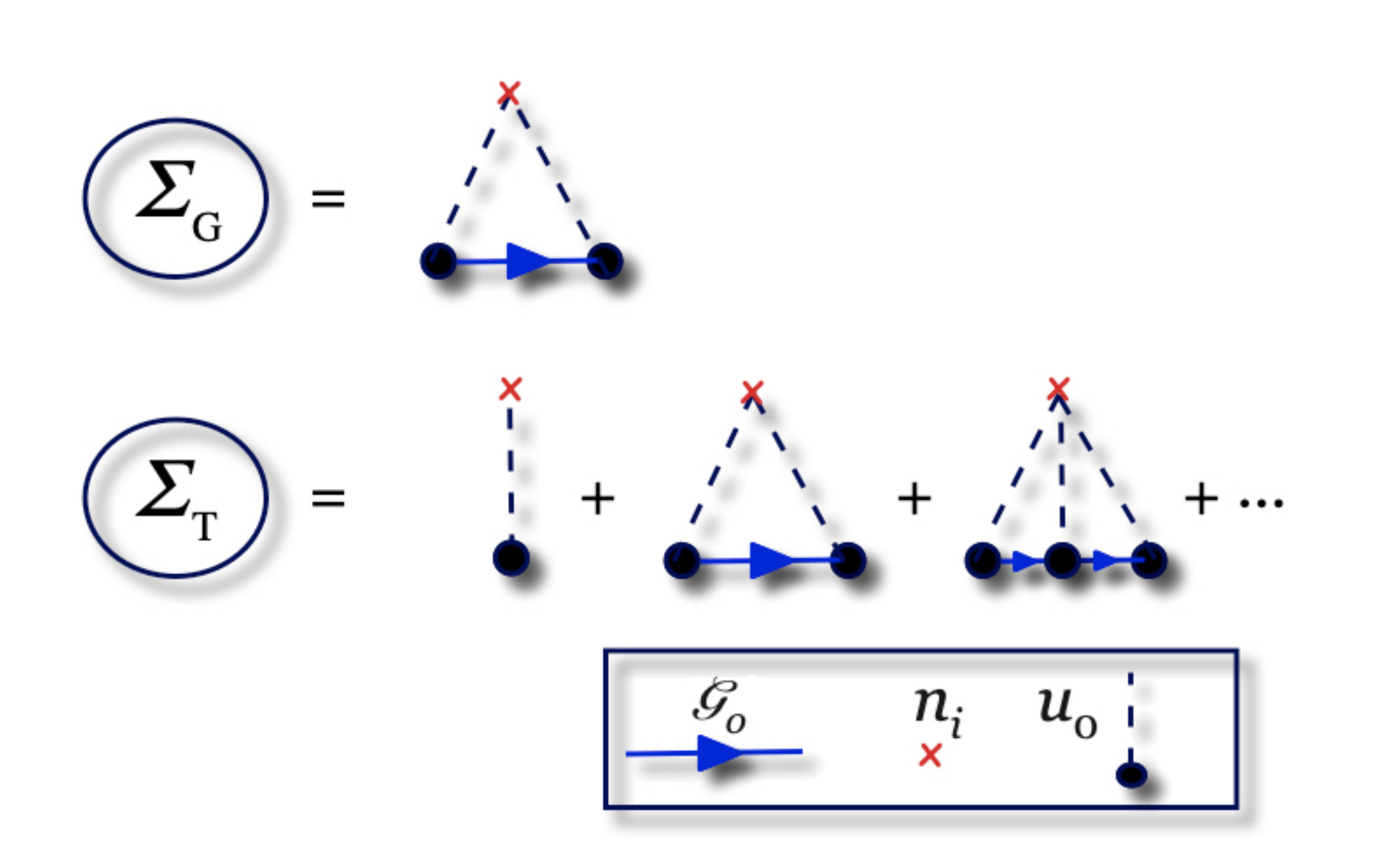} \caption{Disorder self energy in the non-crossing approximation. In the Gaussian
case, the self energy consists of a single ``rainbow'' diagram with
two potential insertions ($\Sigma_{G}$). In the $T$-matrix approach,
one effectively resums the full series of single-impurity scattering
events ($\Sigma_{T}$), which is then proportional to the impurity
areal density $n_{i}$. }
\label{fig:02} 
\end{figure}

In the presence of random impurities with areal density $n_{i}$,
the retarded and advanced Green's functions are dressed with the corresponding
disorder self-energies. The self energy is given by the average over
disorder ($\prec...\succ$) of the $T$-matrix expansion as shown
in Fig. \ref{fig:02}, that is,
\begin{equation}
\Sigma^{R(A)}=T^{R(A)}=n_{i}\frac{u_{0}^{2}\mathcal{G}^{R(A)}}{1-u_{0}\mathcal{G}^{R(A)}}.\label{eq:self_energy_Tmatrix}
\end{equation}
In the above $\mathcal{G}^{R(A)}=G_{0}^{R(A)}(\mathbf{x},\mathbf{x})$
is the local retarded (advanced) clean Green's function, i.e. evaluated
at coinciding space arguments and the disorder average is defined
by 
\begin{eqnarray*}
\prec U\left(\mathbf{x}\right)\succ & = & u_{0}\,,\\
\prec U\left(\mathbf{x}\right)U\left(\mathbf{x'}\right)\succ & = & u_{0}^{2}\delta\left(\mathbf{x}-\mathbf{x}'\right)\,,\\
\prec U\left(\mathbf{x}\right)U\left(\mathbf{x}'\right)U\left(\mathbf{x}''\right)\succ & = & u_{0}^{3}\delta\left(\mathbf{x}-\mathbf{x}'\right)\delta\left(\mathbf{x}'-\mathbf{x}''\right)\,,
\end{eqnarray*}
and so on and so forth.

In the equation of motion, one determines the Green function self-consistently
by replacing $\mathcal{G}^{R(A)}$ with the disorder-average one $G^{R(A)}$.
To quadratic order in the disorder potential, in the so-called first
Born approximation, one keeps only the first term of the series expansion
of Eq. (\ref{eq:self_energy_Tmatrix}). The corresponding retarded/advanced
self-energies read as 

\begin{equation}
\Sigma^{R(A)}=n_{i}u_{0}^{2}\int\frac{d\mathbf{k'}}{(2\pi)^{2}}G_{0\mathbf{k'}}^{R,A}=\mp i\pi n_{i}u_{0}^{2}N_{F}\left\langle P\left(\mathbf{k}_{F}\right)\right\rangle ,\label{eq:self_energy_RA}
\end{equation}
where from \textcolor{black}{now $\left\langle \ldots\right\rangle $
indicates the integration, normalized to $2\pi$, over the wavevector
angle defining }the direction of $\mathbf{k}_{F}$. As a result the
$out-$contribution of the collision integral becomes 
\begin{eqnarray}
I_{out}=-i\left(T^{R}G^{K}-G^{K}T^{A}\right).\label{eq:I_out_expression}\\
\nonumber 
\end{eqnarray}
In a self-consistent evaluation of the self-energy, one should replace
$G_{0\mathbf{k}}^{R,A}$ with the disorder-averaged Green's function
$G_{\mathbf{k}}^{R,A}$ in Eq.(\ref{eq:self_energy_RA}). Provided
one is not too close to the Dirac point, one can neglect the disorder
correction to the electron spectrum and the self-consistent solution
for the self-energy is reasonably approximated by the expression in
Eq.(\ref{eq:self_energy_RA}). 

Next, we consider the Keldysh self-energy, which determines the $in$-contribution
of the collision integral. The perturbation expansion of the Keldysh
Green's function reads quite different from that of the retarded and
advanced Green's functions. After disorder averaging, at each order
in $u_{0}$, there are as many terms as positions in which the Keldysh
component can be placed with the additional requirement that on the
left (right) of the Keldysh component there can be only retarded (advanced)
Green's functions. For instance, at first and second orders one has
\begin{align*}
G^{K\left(1\right)} & =G_{0}^{R}u_{0}G_{0}^{K}+G_{0}^{K}u_{0}G_{0}^{A},\\
G^{K\left(2\right)} & =G_{0}^{R}u_{0}\left(G_{0}^{R}u_{0}G_{0}^{K}+G_{0}^{K}u_{0}G_{0}^{A}\right)+G_{0}^{K}u_{0}G_{0}^{A}u_{0}G_{0}^{A}\,.
\end{align*}
In the end, the Keldysh self-energy then reads

\noindent 
\begin{equation}
\Sigma^{K}=n_{i}u_{0}^{2}\overline{T}^{R}\int\frac{d\mathbf{k}'}{\left(2\pi\right)^{2}}G_{\mathbf{k}'}^{K}\overline{T}^{A},\label{eq:Sigma_K}
\end{equation}
where we have introduced the following notation 
\begin{equation}
T^{R(A)}=n_{i}u_{0}^{2}G^{R\left(A\right)}\overline{T}^{R(A)}\label{eq:T_definition}
\end{equation}
with 
\begin{equation}
\overline{T}^{R(A)}=\frac{1}{1-u_{0}G^{R(A)}}.\label{eq:T_barra_definition}
\end{equation}

\noindent Interestingly, only the $\overline{T}^{R,A}$ parts of the
$T$-matrices appear in the Keldysh self-energy upon performing the
disorder average as a result of the latter involving impurity insertions
in both the retarded and advanced Green's functions at the left and
right of the Keldysh Green's function, respectively. This corresponds
to the so-called $T$-matrix insertion in the vertex correction of
the Kubo linear response formalism, which effectively resums the infinite
set of single-impurity scattering diagrams \citep{Milletari_2016a,Milletari_2016b}.
The ''in'' contribution to the collision integral Eq.(\ref{eq:collision_integral})
finally reads 
\begin{equation}
I_{in}=in_{i}u_{0}^{2}\int\frac{d\mathbf{k}'}{\left(2\pi\right)^{2}}\left(G_{\mathbf{k}}^{R}\overline{T}^{R}G_{\mathbf{k}'}^{K}\overline{T}^{A}-\overline{T}^{R}G_{\mathbf{k}'}^{K}\overline{T}^{A}G_{\mathbf{k}}^{A}\right).\label{eq:I_in_expression}
\end{equation}
After some algebra, it is possible to recast the collision integral
Eq.(\ref{eq:collision_integral}) in the form (for details see Appendix
B) 
\begin{widetext}
\begin{equation}
I=I_{in}+I_{out}=in_{i}u_{0}^{2}\int\frac{d^{2}\mathbf{k'}}{(2\pi)^{2}}\Big(G_{\mathbf{k}}^{R}\overline{T}^{R}G_{\mathbf{k'}}^{K}\overline{T}^{A}-\overline{T}^{R}G_{\mathbf{k'}}^{K}\overline{T}^{A}G_{\mathbf{k}}^{A}-G_{\mathbf{k'}}^{R}\overline{T}^{R}G_{\mathbf{k}}^{K}\overline{T}^{A}+\overline{T}^{R}G_{\mathbf{k}}^{K}\overline{T}^{A}G_{\mathbf{k'}}^{A}\Big).\label{eq:Collision_integral_detailed_balance}
\end{equation}
\end{widetext}

\noindent From the point of view of the kinetic equation, the variables
$\mathbf{k}$ and $\mathbf{k'}$ represent the momentum before and
after a single-impurity scattering event, depending whether one considers
the $in$- and $out$-contributions. One can easily check that the
detailed balance is satisfied. If $\mathbf{k}\leftrightarrow\mathbf{k'}$,
the $in-$ and $out-$terms are interchanged and then the microscopic
reversibility of the scattering probability is preserved \citep{Lebowitz_1995}.
The general expression Eq.(\ref{eq:Collision_integral_detailed_balance})
is one of the main results of this work. The familiar case of a Fermi
gas with Gaussian white-noise correlation is recovered by letting\textcolor{blue}{{}
}\textcolor{black}{$\overline{T}^{R(A)}\rightarrow1$.} 

\section{The Eilenberger equation\label{sec:IV}}

In this section, we solve the kinetic equation presented earlier (cf.
Eq.(\ref{eq:gradient_expansion}) and the following discussion) within
the quasiclassical approximation \citep{Rammer_1986,Rammer_2007,Raimondi_2003}.
The latter has a resemblance to the standard Boltzmann transport theory
that lends itself to a physically transparent interpretation of 2D
spin-orbit coupled transport effects in the experimentally relevant
diffusive regime \citep{Offidani_2018c,Milletari_2016a,Sousa_2020}. 

The established framework in which the quasiclassical theory is expressed
is that of the real-time formulation of the Keldysh technique, as
discussed above. The quasiclassical Green's function $\check{g}$
is usually defined by \citep{Eilenberger_1968}

\begin{equation}
\check{g}\left(\mathbf{x},t,\mathbf{n},\omega\right)\doteq\frac{i}{\pi}\intop_{\mathcal{C}}d\xi\,\check{G}\left(\mathbf{x},t,\mathbf{k},\omega\right)
\end{equation}
where we have introduced the variable $\xi=\epsilon_{1-1}(k)-\varepsilon$
in terms of the dispersion relation of the spin-helical band at the
Fermi energy. Since we assumed $\varepsilon>0$, all the projectors
in what follows are constructed from eigenstates $l=1,n=-1$. The
integration contour $\mathcal{C}$ is taken in such a way to capture
the contribution of the Green's function pole, and that the $\xi$-integration
leaves unaffected the angular dependence described by the unit vector
$\mathbf{n}=\mathbf{k}/\left|\mathbf{k}\right|$. The so-called Eilenberger
equation is then obtained by applying the $\xi$-integration to the
Eq.(\ref{eq:gradient_expansion}) by \textit{reasonably} assuming
that the self-energy does not have a further singular behavior, which
would add to the Green's function pole. At the leading order in the
weak-disorder expansion $\varepsilon\tau\gg1$, with $\tau=1/(2\textrm{Im}\,\Sigma^{A})$
the quasiparticle lifetime, the $\xi$-integration procedure is not
affected by the disorder-induced dressing of the pole. The Eilenberger
equation then reads 
\begin{eqnarray}
\partial_{t}\check{g}+\frac{1}{2}\left\{ \boldsymbol{\sigma\cdot},\left(\boldsymbol{\nabla}-e\mathbf{E\partial_{\omega}}\right)\check{g}\right\} +\qquad\quad\nonumber \\
+i\left[h\left(\mathbf{\mathbf{k}_{F}}\right),\check{g}\right]=-i\left[\check{\Sigma},\check{g}\right].\label{eq:Eilenberger}
\end{eqnarray}
The quasiclassical Green's function has the same triangular matrix
structure of the original Green's function, i.e. 
\begin{equation}
\check{g}=\left(\begin{array}{cc}
g^{R} & g\\
0 & g^{A}
\end{array}\right),
\end{equation}
and in the clean system, the retarded (advanced) quasiclassical Green's
function are 
\begin{equation}
g^{R(A)}=\pm P(\mathbf{k}_{F})\equiv\pm P\left(\theta\right),\label{eq:quasiclassical_greenfunction_R/A}
\end{equation}
where $P(\mathbf{k}_{F})\equiv P\left(\theta\right)$ is the projector
on the spin-helical band evaluated at the Fermi energy with only the
angle dependence remaining. In principle, we have to solve Eq.\,(\ref{eq:Eilenberger})
for all the components of $\check{g}$, but one can considerably simplify
the problem by noting that the elastic scattering from scalar impurities
merely produces a shift in the pole of the retarded (advanced) component.
Hence, as already mentioned, the expression (\ref{eq:quasiclassical_greenfunction_R/A})
remains unchanged at leading order in powers of $\left(\varepsilon\tau\right)^{-1}$.
Ultimately $g^{R(A)}$ gives information about the density of states
whereas $g^{K}\equiv g$ provides the distribution function. Finally,
the Eilenberger equation for the Keldysh component Eq. (\ref{eq:Eilenberger})
becomes
\begin{eqnarray}
\partial_{t}g+\frac{1}{2}\left\{ \boldsymbol{\sigma\cdot},\left(\nabla-e\mathbf{E\partial_{\omega}}\right)g\right\} +i\left[h\left(\mathbf{k_{F}}\right),g\right]=\mathcal{I},\label{eq:Eilenberger_Keldysh}
\end{eqnarray}
where the $\xi$-integration must be performed also in the right hand
side of Eq. (\ref{eq:gradient_expansion}). Explicitly, by applying
the $\xi$-integration to the expression of $I$ given by Eq. (\ref{eq:Collision_integral_detailed_balance})
and anticipating the definition of the scattering time given below
in the Eq. (\ref{eq:scattering_time}), we obtain 
\begin{equation}
\mathcal{I}=-\frac{1}{2\tau_{0}}\left(\left\langle g^{R}\right\rangle \overline{T}^{R}g\overline{T}^{A}-\overline{T}^{R}g\overline{T}^{A}\left\langle g^{A}\right\rangle \right)+\label{eq:T-matrix_coll_integral}
\end{equation}
\[
+\frac{1}{2\tau_{0}}\left(g^{R}\overline{T}^{R}\left\langle g\right\rangle \overline{T}^{A}-\overline{T}^{R}\left\langle g\right\rangle \overline{T}^{A}g^{A}\right),
\]
where, for brevity, we defined $g^{R;A;K}\equiv g^{R,A,K}\left(\theta,\omega\right)$
and its angle averaged $\left\langle g^{R,A,K}\right\rangle =\left\langle g^{R,A;K}\left(\theta',\omega\right)\right\rangle $
with $\theta$ and $\theta'$ the angles of $\mathbf{k}$ and $\mathbf{k'}$,
respectively, both momenta evaluated at the Fermi surface.\textcolor{blue}{{}
}\textcolor{black}{For convenience we have also introduced the basic
scattering rate}

\textcolor{black}{
\begin{equation}
{\color{black}1/\tau_{0}=2\pi N_{0}n_{i}u_{0}^{2}\,,}\label{eq:scattering_time}
\end{equation}
which has the meaning of the inverse quasiparticle lifetime of graphene
without the SOC in the Gaussian limit. }

By looking at Eq. (\ref{eq:Eilenberger_Keldysh}) we note two key
differences with respect to the semiclassical Boltzmann transport
equation (BTE): (1) there is a commutator term $i\left[h\left(\mathbf{k_{F}}\right),g\right]$
missing in the BTEs because there $g$ is a scalar, (2) a general
scattering kernel is present which, in principle, holds for any type
of impurity (magnetic, etc.), whereas semiclassical BTEs based on
a scalar distribution function only applies to scalar impurities.
Although not considered in the present paper, the method can be readily
extended to impurity potential with a matrix structure in the internal
degrees of freedom. 

\noindent In the remainder of this paper, we confine our analysis
to the case of a constant and uniform electric field for which the
space and time derivatives in the Eq. (\ref{eq:Eilenberger_Keldysh})
drop out and the quasiclassical Green's function then depends only
on $\mathbf{n},\omega$ or equivalently on $\theta,\omega$. To leading
order in $\mathbf{E}$, one can safely replace $g$ in the force term
(in the left hand side) by the equilibrium Keldysh Green's function.
The Eilenberger equation (\ref{eq:Eilenberger_Keldysh}) then reduces
to 
\begin{equation}
-\frac{1}{2}\left\{ \boldsymbol{\sigma\cdot},e\mathbf{E\partial_{\omega}}g_{eq}(\theta,\omega)\right\} +i\left[h\left(\mathbf{k_{F}}\right),g\right]=\mathcal{I},\label{eq:linear_response}
\end{equation}
where $g_{eq}(\theta,\omega)=f(\omega)P(\theta)$ is the equilibrium
quasiclassical Keldysh Green's function with

\begin{equation}
f(\omega)=2\tanh\left(\frac{\omega}{2T}\right),
\end{equation}
where $T$ is the temperature (in our natutal units $k_{B}=1$).

To proceed further, following Ref. \citep{Schwab_2011}, we propose
the following ansatz for the quasiclassical Keldysh Green's function

\begin{equation}
g(\theta,\omega)=g_{0}(\theta,\omega)\,P(\theta)\,,\label{eq:g0_definition}
\end{equation}
where $g_{0}\left(\theta,\omega\right)$ is a scalar function. Although
$g$ is still a matrix, its structure is entirely constrained. The
ansatz can be motivated by the following argument. Inspection of the
Eq.(\ref{eq:Eilenberger_Keldysh}) shows that, at leading order in
the dilute regime ($\varepsilon\tau\gg1$), the solution must commute
with the Hamiltonian and be of order $g\sim\tau\sim\tau_{0}$. The
commutator in the left hand side, although of order $\varepsilon\tau\gg1$,
vanishes because the Hamiltonian density at the Fermi level can be
written as $h\left(\mathbf{k_{F}}\right)=\varepsilon P(\theta)$ and
the remaining terms in the equation are of order unity. We note that
this ansatz may not be sufficient when one is dealing with quantum
(weak-localization) corrections in the weak-disorder expansion. Using
these ingredients and by taking the trace of the Eq.(\ref{eq:T-matrix_coll_integral})
one gets a scalar collision integral for $g_{0}$

\begin{equation}
\mathcal{I}_{0}=-\int_{0}^{2\pi}\frac{\mathrm{d\theta'}}{2\pi}W\left(\vartheta\right)\left[g_{0}\left(\theta,\omega\right)-g_{0}\left(\theta',\omega\right)\right],
\end{equation}
where

\noindent 
\begin{eqnarray}
W\left(\vartheta\right)=\frac{N_{F}}{N_{0}}\frac{1}{\tau_{0}}\mathrm{Tr}\left[P(\theta)\overline{T}^{R}P(\theta^{\prime})\overline{T}^{A}\right]\label{eq:total_Kernel_potential}
\end{eqnarray}

\noindent is a function of the angle difference that from now we call
$\vartheta\equiv\theta-\theta'$. 

In the end, once the solution for the quasiclassical Green's function
$g_{0}(\theta,\omega)$ is known one can easily obtain the steady-state
observables, such as the electric current, the spin polarization and
the spin current densities. According with the general recipe in the
Abelian case \citep{Konschelle_2014,Kadanoff_1962}, using the notation
\citep{Rammer_1986}

\begin{equation}
\intop\frac{d\mathbf{k}}{(2\pi)^{2}}\int\frac{d\omega}{2\pi i}\approx N_{F}\int\frac{d\mathbf{n}}{2\pi}\int\frac{d\omega}{2\pi i}\int d\xi,
\end{equation}

\noindent if $O=\sigma_{i}\otimes s_{j}$ indicates a generic observable,
its quantum statistical non-equilibrium average $\mathcal{O}$ is
given by 
\begin{eqnarray}
\mathcal{O} & \equiv & \frac{i}{2}\intop\frac{d\omega}{2\pi}\intop\frac{d\mathbf{k}}{(2\pi)^{2}}\:\mathrm{Tr}\left[OG^{K}\left(\mathbf{k},\omega\right)\right]\nonumber \\
 & = & -\frac{N_{F}}{4}\intop d\omega\langle Tr\,\left[Og\left(\theta,\omega\right)\right]\rangle\nonumber \\
 & = & -\frac{N_{F}}{4}\intop d\omega\langle g_{0}\left(\theta,\omega\right)Tr\,\left[OP\left(\theta\right)\right]\rangle,\nonumber \\
 & = & -\frac{N_{F}}{2}\intop d\omega\langle g_{0}\left(\theta,\omega\right)O_{{\bf k}}\rangle,\label{eq:observables}
\end{eqnarray}
where $O_{\mathbf{k}}=\textrm{Tr}\left[O\,P(\theta)\right]/2$ indicates
the equilibrium average and, as before, $\left\langle \ldots\right\rangle $
is the integration over the directions of $\mathbf{k}=k_{F}\mathbf{n}$.
The trace symbol involves all internal degrees of freedom: sublattice,
valley and spin. Because intervalley scattering is neglected for simplicity
(the impurity potential $U(\mathbf{x})$ is a scalar quantity), the
trace over the valley degree of freedom yields a simple factor of
two, which is compensated by the factor of two in the denominator
in the definition of the equilibrium average of $O_{\mathbf{k}}$.
The relevant observables are the charge current \textcolor{black}{($\mathbf{J}=-ev\boldsymbol{\mathbf{\sigma}}\otimes s_{0}$)},
spin polarization ($\mathbf{S}=\frac{\hbar}{2}\sigma_{0}\otimes\mathbf{s}$)
and z-polarized spin Hall current ($\mathbf{J}_{s}^{z}=\frac{\hbar v}{2}\boldsymbol{\sigma}\otimes s_{z}$).
Here we have reinstated the Planck constant for the sake of clarity.
For the $C_{3v}$-invariant model with spin-valley coupling, the equilibrium
charge current, spin polarization density and (persistent) spin Hall
current can be easily evaluated as
\begin{equation}
\mathbf{J}_{\mathbf{k}}=(-ev)\frac{2(\beta+\alpha\gamma)}{1+\alpha^{2}+\beta^{2}+\gamma^{2}}\mathbf{n}\,,\label{eq:s1}
\end{equation}

\begin{center}
\begin{equation}
\mathbf{S}_{\mathbf{k}}=\hbar\frac{(\alpha+\beta\gamma)}{1+\alpha^{2}+\beta^{2}+\gamma^{2}}\hat{z}\times\mathbf{n}\,,\label{eq:s2}
\end{equation}
\par\end{center}

\begin{equation}
\mathbf{J}_{s\mathbf{k}}^{z}=(\hbar v)\frac{(\beta-\alpha\gamma)}{1+\alpha^{2}+\beta^{2}+\gamma^{2}}\mathbf{n}\,,\label{eq:s3}
\end{equation}
where $\alpha,\beta$ and $\gamma$ depend on the parameters of the
Hamiltonian and on the Fermi energy as defined in the Eq.(\ref{eq:generic_eigenvector}). 

Similar to the well-studied high density regime with two occupied
Dirac bands \citep{Milletari_2017,Offidani_2017,Offidani_2018b,Sousa_2020},
the standard first Born approximation will allows us to obtain the
leading semiclassical contribution to the nonequilibrium spin polarization
density generated by the inverse spin galvanic effect, namely $\mathcal{S}_{i}\propto\epsilon_{ij}\epsilon\tau\,E_{j}$,
with $\epsilon_{ij}$ the Levi-Civita symbol and $i,j=x,y$. On the
other hand, the calculation of the steady-state $\hat{z}$-polarized
spin Hall current density, $\mathcal{J}_{i}^{z}\propto\epsilon_{ij}\epsilon\tau\,E_{j}$,
will require higher-order corrections to the self-energy due to the
pivotal role played by skew scattering mechanism in the extrinsic
spin Hall effect.

\section{The Eilenberger equation in the First Born approximation\label{sec:V}}

In this section we consider the self-consistent Born approximation,
which amounts to confine to second order in the disorder potential
expansion, implying \textcolor{black}{$\overline{T}^{R(A)}\longrightarrow1$.}
As a result, the matrix transport equation (\ref{eq:linear_response})
reduces to a simpler scalar effective transport equation for the ``charge''
component $g_{0}\left(\theta,\omega\right)$ of the quasiclassical
Green's function 
\begin{equation}
\left(\mathbf{E}\cdot\mathbf{J}_{{\bf k}}\right)\partial_{\omega}f=-\frac{1}{2\pi}\intop_{0}^{2\pi}d\theta'\,W\left(\vartheta\right)\left(g_{0}\left(\theta,\omega\right)-g_{0}\left(\theta',\omega\right)\right),\label{eq:Transport_Equation}
\end{equation}
where 
\begin{equation}
W\left(\vartheta\right)=\frac{N_{F}}{N_{0}}\frac{1}{\tau_{0}}\mathrm{Tr}\{P\left(\theta\right),P\left(\theta'\right)\}.\label{eq:effective_kernel}
\end{equation}
is the scattering kernel in the first Born approximation {[}c.f. Eq.
(\ref{eq:total_Kernel_potential}){]}. We remind the reader that,
unless otherwise stated, the projectors are evaluated at the Fermi
surface. Due to the periodicity of the projectors, the scattering
kernel can only depend on the difference of the two angles. Furthermore
due to the cyclic property of the trace, the expression is invariant
under the interchange of the two angles, implying that the effective
kernel is an \textit{even periodic function} of the angle difference
$\vartheta\equiv\theta-\theta'$. The transport equation Eq.(\ref{eq:Transport_Equation})
together with the expressions of the effective kernel Eq.(\ref{eq:effective_kernel})
and the velocity vertex Eq.(\ref{eq:s1}), defines the coupled spin-charge
response of the projected band at the Fermi energy. This is one of
the central results of this paper and can be used to obtain the charge
conductivity and spin-galvanic conductivity of the spin-helical band
as discussed below.

We first illustrate the formalism with the $C_{6v}$-invariant Dirac-Rashba
model obtained by setting $\lambda_{\textrm{sv}}=0$ in Eq. (\ref{eq:SOC});
the corresponding band structure and spin texture is shown in Fig.
\ref{fig1}(a). In this case, the eigenvalue coefficients are $\alpha=-\beta=-\varepsilon/\sqrt{\varepsilon(\varepsilon+2\lambda)}$,
$\gamma=-1$ and one can obtain simple analytical expressions for
all the transport coefficients. In particular the equilibrium averages
(\ref{eq:s1}-\ref{eq:s3}) read as
\begin{eqnarray}
{\bf J}_{{\bf k}} & = & (-ev)\frac{\sqrt{\varepsilon(\varepsilon+2\lambda)}}{\varepsilon+\lambda}{\bf n}\label{eq:s1_0}\\
{\bf S}_{{\bf k}} & = & -\hbar\frac{1}{2}\frac{\sqrt{\varepsilon(\varepsilon+2\lambda)}}{\varepsilon+\lambda}{\hat{{\bf z}}}\times{\bf n}\label{eq:s2_0}\\
\mathbf{J}_{s\mathbf{k}}^{z} & = & 0.\label{eq_s3_0}
\end{eqnarray}
The scattering kernel within the first Born approximation reads as
(see Appendix \ref{details_W} for details) 
\begin{equation}
W(\vartheta)=\frac{1}{\tau_{0}}\frac{((2\lambda+\varepsilon)\cos(\theta-\theta')+\varepsilon)^{2}}{\varepsilon(\lambda+\varepsilon)}\,.\label{eq:Rashba_only_kernel}
\end{equation}
One sees that the scattering kernel is left-right symmetric ($W(\vartheta)=W(-\vartheta)$)
to all orders in the scattering potential, i.e. skew-scattering is
absent and no extrinsic spin Hall effect (SHE) can be expected in
this case. Interestingly, the absence of any form of SHE is an exact
property of the model provided that the impurities are non-magnetic.
This is a consequence of the in-plane spin texture of the minimal
2D Dirac-Rashba model and can also be understood from the exact Ward
identities for the 4-point vertex function \citep{Milletari_2017}.

The solution of the transport equation (\ref{eq:Transport_Equation})
reads 
\begin{equation}
g_{0}(\theta,\omega)=-(\partial_{\omega}f)\,\tau_{s}\,{\bf J}_{{\bf k}}\cdot{\bf E},\label{eq:solution_Rashba_only_first}
\end{equation}
where 
\begin{eqnarray}
\frac{1}{\tau_{s}} & = & \frac{1}{2\pi}\int_{0}^{2\pi}{\rm d}\theta W(\theta)(1-\cos(\theta))\nonumber \\
 & = & \frac{1}{2\tau_{0}}\frac{\varepsilon^{2}+4\lambda^{2}}{\varepsilon(\varepsilon+\lambda)}.\label{eq:transport_time_Rashba_only}
\end{eqnarray}
By using the explicit expressions of the transport time (\ref{eq:transport_time_Rashba_only})
and of the equilibrium average (\ref{eq:s1_0}) one obtains 
\begin{equation}
g_{0}(\theta,\omega)=(ev)(2\tau_{0})(\partial_{\omega}f)\frac{\varepsilon\sqrt{\varepsilon(2\lambda+\varepsilon)}}{4\lambda^{2}+\varepsilon^{2}}\,\mathbf{E}\cdot\mathbf{n},\label{eq:solution_Rashba_only}
\end{equation}
By inserting this last result in the Eq.(\ref{eq:observables}) and
considering the equilibrium average (\ref{eq:s2_0}) one gets the
following current-induced spin polarization 
\begin{equation}
\boldsymbol{\mathcal{S}}=\frac{ev\tau_{0}}{\pi v^{2}}\frac{\varepsilon^{2}(\varepsilon+2\lambda)}{(\varepsilon^{2}+4\lambda^{2})}\hat{\mathbf{z}}\times\mathbf{E},\label{eq:Edelstein_Rashba_only}
\end{equation}
while the charge current reads 
\begin{equation}
\boldsymbol{\mathcal{J}}=\frac{2e^{2}\tau_{0}}{\pi\hbar}\frac{\varepsilon^{2}(2\lambda+\varepsilon)}{(4\lambda^{2}+\varepsilon^{2})}\mathbf{E}\,.\label{eq:Currentx_Rashba_only}
\end{equation}
These relations imply 
\begin{equation}
\boldsymbol{\mathcal{S}}=\frac{\hbar}{2ev}\,\hat{\mathbf{z}}\times\boldsymbol{\mathcal{J}}\,.\label{eq:locking}
\end{equation}

The current-induced spin polarization is in plane and orthogonal to
the applied electric field as a consequence of the symmetries of the
model. In fact, it is easy to see that the inversion symmetry in the
plane of graphene implies $\mathcal{S}_{z}=0$ and $\boldsymbol{\mathcal{S}}\cdot\mathbf{E}=0$.
This is also the case in the presence of sublattice-staggered interactions
(i.e. $\{\lambda_{\textrm{sv}},\Delta\}\neq0$) due to the existence
of a mirror reflection symmetry in the plane of the heterostructure
\citep{Sousa_2020}. The locking of non-equilibrium spin polarization
density and charge current in plane at 90 degrees Eq. (\ref{eq:locking})
is thus a general property of non-magnetic graphene heterostructures
with $C_{6v}$ or $C_{3v}$ point group symmetry. These restrictions
are lifted in magnetic graphene heterostructures, where $\boldsymbol{\mathcal{S}}$
acquires collinear and out of plane components \citep{Sousa_2020}.
We note that Eqs.(\ref{eq:Edelstein_Rashba_only})-(\ref{eq:Currentx_Rashba_only})
coincide exactly with the results for the electric-field-induced spin
polarization \citep{Edelstein_1990} obtained via the Kubo-Streda
linear response theory by Offidani \emph{et al} \citep{Offidani_2017}
and confirm the equivalence of the two approaches.

\section{The Skew-Scattering mechanism\label{sec:VI}}

We now consider the self-energy expansion beyond the first Born approximation.
This is relevant for models with non-coplanar spin texture ($\lambda_{\textrm{sv}}\neq0$
and $\lambda\neq0$), for which skew scattering mechanism is active
and thus the model supports an extrinsic SHE with a semiclassical
scaling $J_{i}^{z}\propto\epsilon_{ij}\epsilon\tau E_{j}$, in addition
to the intrinsic SHE driven by Berry curvature effects \citep{Offidani_2018c,Milletari_2017,Garcia_2017}.
The extrinsic mechanism is expected to provide the dominant contribution
to the SHE in ultra-clean devices with high charge carrier mobility
(i.e. $\epsilon\tau\gg1$). When the scattering potential is not too
strong ($|{\cal G}^{R\left(A\right)}|u_{0}\ll1$), one may expand
the matrices $\overline{T}^{R(A)}$ {[}c.f. Eqs.(\ref{eq:T_barra_definition}){]}
as 
\begin{equation}
\overline{T}^{R(A)}\simeq1+u_{0}\textrm{Re}{\cal G}^{R(A)}\mp iu_{0}\textrm{Im}{\cal G}^{R(A)}.\label{eq:Tmatrix_expansion}
\end{equation}
This approximation corresponds to keeping the first two diagrams in
the second line of Fig.\,\,\ref{fig:02}. We stress that the real
part in Eq. (\ref{eq:Tmatrix_expansion}) does not give contribution
to the skew scattering mechanism, but only renormalises the lowest-order
scattering amplitude. Mathematically, the \textit{skewness} in the
effective scattering kernel $W(\vartheta)\neq W(-\vartheta)$ results
from the imaginary part of the retarded/advanced $T$-matrix, which
endows the scattering term in Eq. (\ref{eq:Transport_Equation}) with
an asymmetric contribution, 
\begin{equation}
W\left(\vartheta\right)=W(\vartheta)+W_{\textrm{ss}}(\vartheta)
\end{equation}
with 
\begin{eqnarray}
W_{\textrm{ss}}\left(\vartheta\right)=-ig_{ss}\left(\frac{N_{F}}{N_{0}}\right)^{2}\frac{2}{\tau_{0}}\mathrm{Tr}\left\langle P\left(\theta''\right)\right\rangle \left[P\left(\theta\right),P\left(\theta'\right)\right]\,,\label{eq:skew_scattering_kernel}
\end{eqnarray}
where the magnitude of the effect is controlled by the ``coupling
constant'' 
\begin{equation}
g_{ss}=2\pi u_{0}N_{0}.\label{eq:gss_coupling_parametrization}
\end{equation}
The commutator under the trace implies that $W_{\textrm{ss}}$ is
odd upon the interchange of $\theta$ and $\theta'$ (in particular,
$W_{\textrm{ss}}$ vanishes for $\theta=\theta^{\prime}$). A shift
$\theta+\psi$, $\theta'+\psi$ clearly leaves $W_{\textrm{ss}}$
unchanged because of the periodicity of the projectors with respect
to both angles, and hence there can be no dependence on the sum $\theta+\theta'$.
As a result, $W_{\textrm{ss}}$ must be an o\textit{dd periodic function}
of $\vartheta=\theta-\theta'$.

The solution of the transport equation generalizes Eq.(\ref{eq:solution_Rashba_only_first})
and reads as 
\begin{equation}
g_{0}(\theta,\omega)=-(\partial_{\omega}f)\left(\tau_{\parallel}\mathbf{E}\cdot\mathbf{J}_{\mathbf{k}}+\tau_{\perp}\mathbf{E}\times\mathbf{\mathbf{J}_{\mathbf{k}}}\cdot{\hat{{\bf z}}}\right)\,,\label{eq:geberal_soution}
\end{equation}
where 
\begin{align}
\tau_{\parallel} & =\frac{\tau_{s}}{1+\left(\frac{\tau_{s}}{\tau_{a}}\right)^{2}}\,,\quad\tau_{\perp}=\frac{\tau_{a}}{1+\left(\frac{\tau_{a}}{\tau_{s}}\right)^{2}}\,\label{eq:transport_times}
\end{align}
are transport times defined through the microscopic rates 
\begin{equation}
\tau_{s}^{-1}=\frac{1}{2\pi}\intop_{0}^{2\pi}d\vartheta\,W(\vartheta)(1-\cos(\vartheta))\label{eq:tau_par}
\end{equation}
\begin{equation}
\tau_{a}^{-1}=\frac{1}{2\pi}\intop_{0}^{2\pi}d\vartheta\,W(\vartheta)\sin(\vartheta).\label{eq:tau_perp}
\end{equation}
These rates generalize the well-known transport rates in semiclassical
transport theory to include the finite asymmetric rates generated
by spin-orbit scattering. The expression is formally equivalent to
the solution of the linearized Boltzmann transport equations for 2D
fermions with SOC \citep{Milletari_2016a,Offidani_2018c}. In Appendix
\ref{traces_expressions} we derive the expression of the microscopic
rates $\tau_{s}^{-1}$ and $\tau_{a}^{-1}$ in terms of the parameters
$\alpha$, $\beta$ and $\gamma$ defining the generic eigenvector
(\ref{eq:generic_eigenvector}) 
\begin{eqnarray}
\frac{\tau}{\tau_{s}} & = & \frac{2N_{F}}{N_{0}}\frac{\gamma^{4}+1+(\alpha^{2}+\beta^{2})(\alpha^{2}+\beta^{2}-\gamma^{2}-1)}{(1+\alpha^{2}+\beta^{2}+\gamma^{2})^{2}},\label{tau_par_explicit}\\
\frac{\tau}{\tau_{a}} & = & 2g_{ss}\left(\frac{N_{F}}{N_{0}}\right)^{2}\frac{\left(1-\gamma^{2}\right)\left(\alpha^{2}+\beta^{2}\right)\left(\alpha^{2}+\beta^{2}-\gamma^{2}-1\right)}{(1+\alpha^{2}+\beta^{2}+\gamma^{2})^{3}}.\label{tau_perp_explicit}
\end{eqnarray}

The zero-temperature nonequilibrium averages to be discussed below
are computed according to Eq.(\ref{eq:observables}), taking into
account the solution (\ref{eq:geberal_soution}). After the integration
over $\omega$ one obtains 
\begin{equation}
g_{0}\left(\theta\right)\equiv\int_{-\infty}^{\infty}d\omega g_{0}\left(\theta,\omega\right)=-4\left(\tau_{\parallel}\mathbf{E}\cdot\mathbf{J}_{\mathbf{k}}+\tau_{\perp}\mathbf{E}\times\mathbf{\mathbf{J}_{\mathbf{k}}}\cdot{\hat{{\bf z}}}\right)\,.\label{eq:omega_integrated_g0}
\end{equation}
As a result the nonequilibrium average of a generic observable reads
\begin{eqnarray}
\mathcal{O} & = & -\frac{N_{F}}{2}\left\langle O_{\mathbf{k}}g_{0}\left(\theta\right)\right\rangle \nonumber \\
 & = & 2N_{F}\left(\tau_{\parallel}\langle O_{\mathbf{k}}\mathbf{E}\cdot\mathbf{J}_{\mathbf{k}}\rangle+\tau_{\perp}\langle O_{\mathbf{k}}\mathbf{E}\times\mathbf{\mathbf{J}_{\mathbf{k}}}\cdot{\hat{{\bf z}}}\rangle\right).\label{eq:av}
\end{eqnarray}

Finally, from the expression Eq.(\ref{eq:av}) for the physical observables,
the Drude conductivity ($\sigma_{xx}=\sigma_{yy}=\mathcal{J}_{x}/E_{x}$),
the spin galvanic susceptibility ($\chi_{yx}=-\chi_{xy}=\mathcal{S}_{y}/E_{x}$)
and the spin Hall conductivity $(\sigma_{\textrm{sH}}\equiv\sigma_{yx}^{z}=-\sigma_{xy}^{z}=\mathcal{J}_{sy}^{z}/E_{x})$
read respectively\textcolor{black}{{} :}
\begin{eqnarray}
\sigma_{xx} & = & \sigma_{0}\frac{|\varepsilon|\tau_{\parallel}}{\hbar}\frac{N_{F}}{N_{0}}\frac{4(\beta+\alpha\gamma)^{2}}{(1+\alpha^{2}+\beta^{2}+\gamma^{2})^{2}}\,,\label{eq:Drude_eff_fin}\\
\chi_{yx} & = & -\chi_{0}\frac{\varepsilon\tau_{\parallel}}{\hbar}\frac{N_{F}}{N_{0}}\frac{(\alpha+\beta\gamma)(\beta+\alpha\gamma)}{(1+\alpha^{2}+\beta^{2}+\gamma^{2})^{2}}\,,\label{eq:Edelstein_eff_fin}\\
\sigma_{yx}^{z} & = & \sigma_{\textrm{sH}}\frac{\varepsilon\tau_{\perp}}{\hbar}\frac{N_{F}}{N_{0}}\frac{(\beta-\alpha\gamma)(\beta+\alpha\gamma)}{(1+\alpha^{2}+\beta^{2}+\gamma^{2})^{2}}\,,\label{eq:SpinHall_eff_fin}
\end{eqnarray}
with $N_{F}$ given in Eq.(\ref{eq:dos_3}) and
\begin{align}
\sigma_{0} & \equiv\frac{2e^{2}}{h}\,,\quad\chi_{0}\equiv\frac{e}{\pi v}\,,\quad\sigma_{\textrm{sH}}\equiv\frac{e}{\pi}\,.\label{eq:constants}
\end{align}
From the above equations, one sees that $\sigma_{xx}$, $(ev/\hbar)\chi_{yx}$
and $(e/\hbar)\sigma_{yx}^{z}$ are expressed in units of conductance
quantum. With this choice, the expressions (\ref{eq:Drude_eff_fin}-\ref{eq:SpinHall_eff_fin})
depend only on dimensionless combinations of the various parameters.
Following Ref. \citep{Offidani_2017}, we quantify the charge-to-spin
conversion efficiency (CSC) with the the following figure of merit
($v_{F}=(1/2\pi)(k_{F}/N_{F})$)
\begin{equation}
CSC=(2ev_{F}/\hbar)\chi_{yx}/\sigma_{xx}.\label{CSC_eff}
\end{equation}
\begin{figure}
\includegraphics[width=80mm]{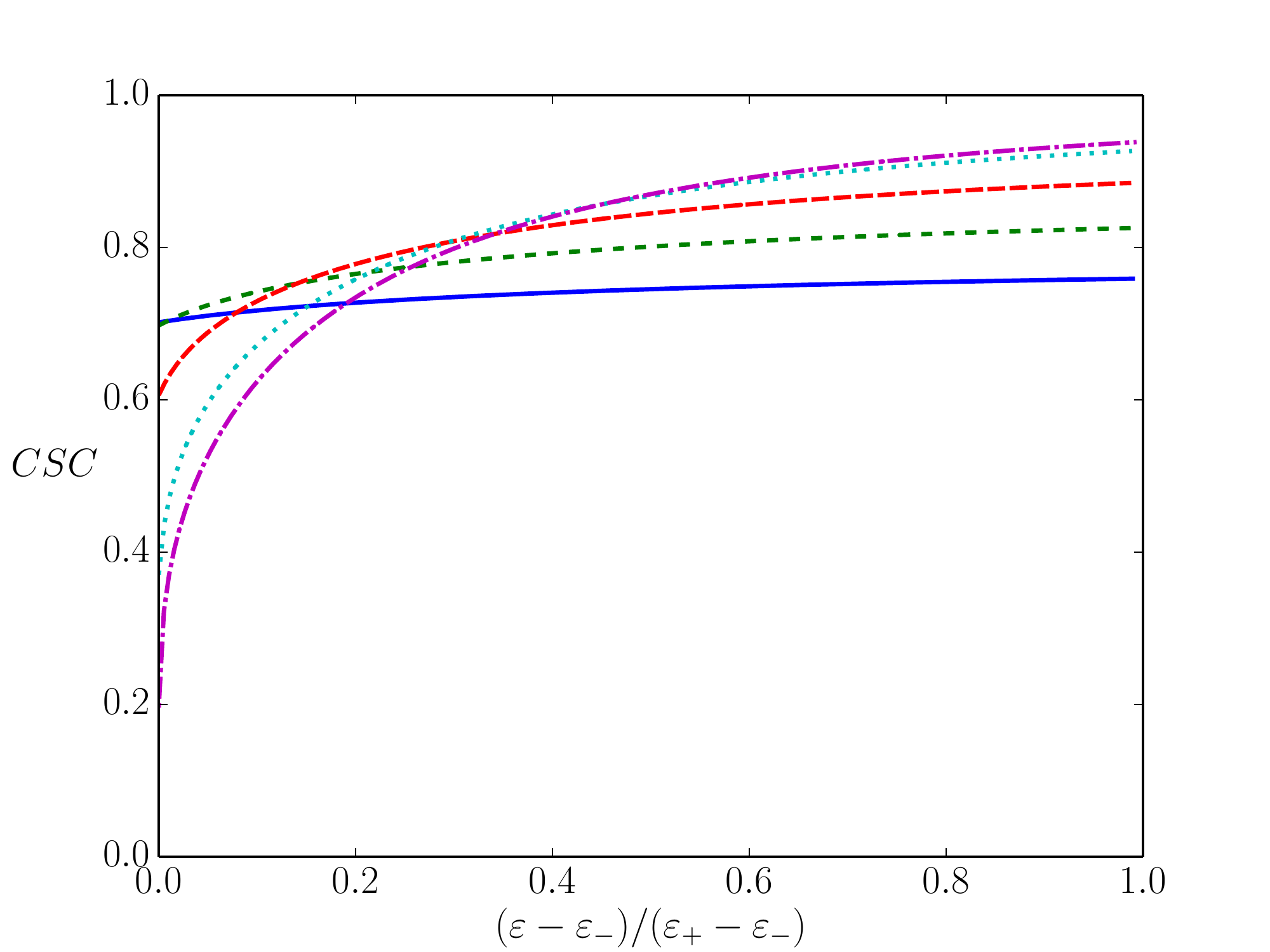} \caption{CSC efficiency as function of the difference of Fermi energy with
respect to the lower edge $\varepsilon_{-}$ of the spin-helical regime,
normalized to the energy interval $\varepsilon_{+}-\varepsilon_{-}$
of the same regime. 
The $T$-matrix expansion parameter, relevant for the skewness of
the potential, is fixed to $(\lambda/\varepsilon)g_{ss}=0.1$. All
energies are measured in units of the Rashba SOC $\lambda$. The various
curves are for different values of the spin-valley coupling: $\lambda_{sv}=0.1\lambda$
(dot-dashes, magenta), $\lambda_{sv}=0.2\lambda$ (dotted, light blue),
$\lambda_{sv}=0.4\lambda$ (long-dashes, red), $\lambda_{sv}=0.6\lambda$
(short-dashes, green), $\lambda_{sv}=0.8\lambda$ (full, blue)}
\label{fig:03} 
\end{figure}

\begin{figure}
\includegraphics[width=80mm]{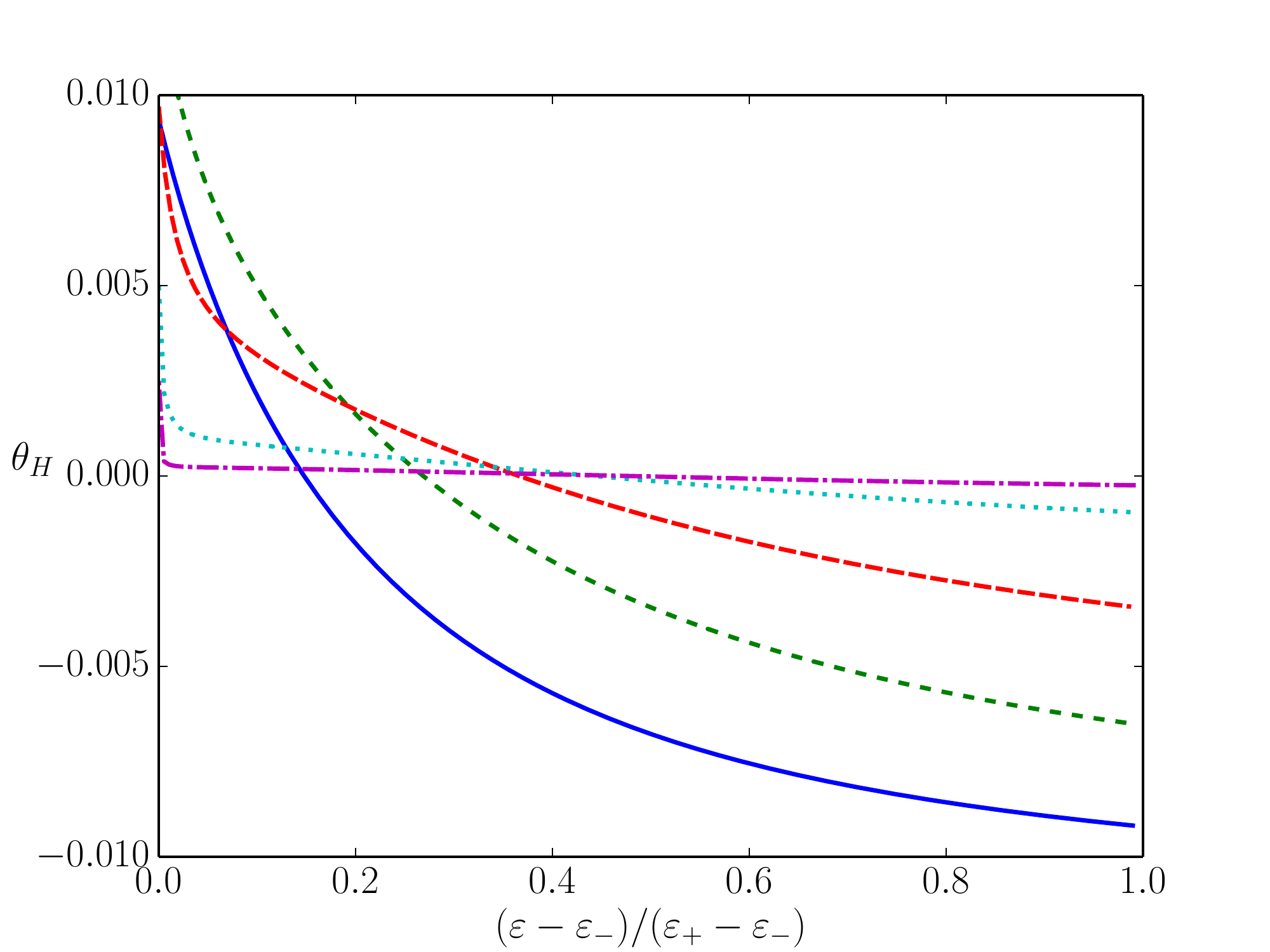} \caption{Spin Hall angle as function of the difference of Fermi energy with
respect to the lower edge $\varepsilon_{-}$ of the spin-helical regime,
normalized to the energy interval $\varepsilon_{+}-\varepsilon_{-}$
of the same regime. All parameters as in fig. \ref{fig:03}. The vanishing
at $\varepsilon=\lambda$, irrespective of the value of the spin-valley
coupling, occurs at different values in terms of normalized variable
$(\varepsilon-\varepsilon_{-})/(\varepsilon_{+}-\varepsilon_{-})$
and precisely at 0.473, 0.442, 0.366, 0.269, 0.147 for $\lambda_{sv}=$0.1,
0.2, 0.4, 0.6, 0.8, respectively.}
\label{fig:04} 
\end{figure}
For the SHE, we introduce similarly the spin Hall angle (SHA) $\theta_{H}$
\begin{equation}
\theta_{H}=(e/\hbar)\sigma_{yx}^{z}/\sigma_{xx}.\label{SHA}
\end{equation}
In the numerical analysis to be presented below, we will express all
energies in units of $\lambda$. The disorder enters through the standard
combination $\varepsilon\tau/\hbar$ and the coupling constant $g_{ss}$
defined in Eq.(\ref{eq:gss_coupling_parametrization}). For a more
extended discussion see Appendix \ref{disorder_estimates}. The equations
(\ref{eq:Drude_eff_fin}-\ref{eq:SpinHall_eff_fin}) clearly show
the different role played by the effective vertices and transport
times in determining the behavior of the physical observables. The
effective vertices depend only on the sublattice-spin entangled nature
of the eigenstate (\ref{eq:generic_eigenvector}). The analytic expression
of the transport times $\tau_{\parallel}$ and $\tau_{\perp}$, which
can be obtained by inserting Eqs. (\ref{eq:alpha}-\ref{eq:gamma})
in Eqs.(\ref{tau_par_explicit}-\ref{tau_perp_explicit}), is too
cumbersome to be presented here, although their numerical evaluation
is straightforward.

The skew-scattering rate $\tau_{a}^{-1}$ shows a characteristic sign
change upon increasing the Fermi energy. Such a sign change can be
recognized by looking at the analytical expression (\ref{tau_perp_explicit}).
When the Fermi energy matches exactly the BR energy scale, i.e. $|\varepsilon|=|\lambda|$,
the coefficients of the eigenvector (\ref{eq:generic_eigenvector})
acquire a simple expression 
\begin{equation}
\alpha=-\frac{\lambda+\lambda_{sv}}{\sqrt{3\lambda^{2}+\lambda_{sv}^{2}}},\beta=\frac{\lambda-\lambda_{sv}}{\sqrt{3\lambda^{2}+\lambda_{sv}^{2}}},\gamma=-1,\label{sign_change_point}
\end{equation}
which implies the vanishing of the factor $1-\gamma^{2}$, in the
formula (\ref{tau_perp_explicit}) for the microscopic rate $\tau_{a}^{-1}$.
Hence, remarkably, the sign change occurs at a well defined energy
when the \textcolor{black}{structure of the eigenvector implies the
vanishing of the skew scattering already in lowest order for the effective
scattering amplitude in the eigenstates (cf. Eq.(\ref{eq:scattering_amplitude_eigenstate})
and the discussion in Appendix \ref{details_W})}\textcolor{blue}{.}
In this respect it is illuminating to notice the following equilibrium
average 
\begin{equation}
\langle\Phi(k_{F},\theta)|\sigma_{0}s_{z}+\sigma_{z}s_{0}|\Phi(k_{F},\theta)\rangle=2(1-\gamma^{2}),\label{remarkable_equation}
\end{equation}
i.e. the factor controlling the sign change, is the equilibrium average
\textcolor{black}{of the z-axis component of the total internal angular
momentum, which includes both spin and pseudo spin. A change of sign
as function of the Fermi energy was also observed for the anomalous
Hall effect of Dirac fermions in \citep{Offidani_2018c}, where it
was interpreted in terms of the behavior of the equilibrium spin texture
of eigenstates. }

Figure \ref{fig:03} shows the Fermi energy dependence of the CSC
efficiency. It is apparent that the figure of merit increases monotonically
with the carrier density within the spin-helical regime, attaining
a maximum at $|\varepsilon|=\varepsilon_{+}$. In regime II, the CSC
efficiency decreases monotonically with the Fermi energy $|\varepsilon|$
\citep{Offidani_2017} due to the presence of two bands with opposite
spin texture. Hence the upper edge $\varepsilon=\pm\varepsilon_{+}$
(for electrons/holes) sets the Fermi energy at which the inverse spin
galvanic effect is the most efficient. Upon increasing the spin-valley
coupling, there is an interesting effect depending on the charge carrier
density. Near the edge of the \textcolor{black}{spin-minority} band
($\varepsilon_{+}$), the spin-valley coupling reduces the CSC efficiency,
with an opposite effect observed near the lower edge where the ``Mexican
hat'' feature develops (see Fig. \ref{fig1} (c)). Such a behavior
correlates well with the two different-sign regimes for the SHA (see
Fig. \ref{fig:04}). The latter changes from positive to negative
upon increasing the Fermi energy $|\varepsilon|$, which is a consequence
of the sign change in the microscopic scattering rate $\tau_{a}^{-1}$
previously discussed and occurs when the Fermi energy matches the
Rashba SOC. 

In the spin precession measurements reported in Ref.\textcolor{black}{{}
\citep{Benitez_2020}, the spin galvanic nonlocal resistance ($R_{\textrm{SGE}}$)
in a graphene/WS$_{2}$ lateral device indeed shows a maximum, in
absolute value, as function of the back-gate voltage ($V_{g}$) for
both charge carrier polarities. Considering that the charge-neutrality
point in the experiment is approximately located at $V_{g}\approx-10$\,V
and the nonlocal signal maximum (for electrons) at $V_{g}\approx-5$\,V,
one may estimate a charge carrier density $n\approx3.5\times10^{11}\:\mathrm{cm}^{-2}$
at the upper edge of the spin-helical regime (see, for example, Ref.
\citep{Ferreira11}). Using the relation between electronic density
and Fermi wavevector for regime I ($n=k_{F}^{2}/2\pi$), we find $\varepsilon_{+}\approx60$
meV. By reasonably assuming $\lambda_{\textrm{sv}}\ll\lambda$ (in
accord with a recent study \citep{Pezo_2020}), one can further estimate
$\lambda\approx29$ meV. This is a somewhat suprising result, given
that functional theory calculations predict a proximity-induced BR
SOC on the order of 0.4 meV \citep{Pezo_2020,Gmitra_2016}. In contrast,
our estimate compares well with recent}\textcolor{blue}{{} }quantum
Hall effect measurements indicating $\lambda\approx15$ meV in graphene/WSe$_{2}$\textcolor{black}{{}
\citep{Wang_21}. Strictly speaking, our estimate should be understood
as an upper bound to the true BR SOC energy scale since the gate voltage
dependence of $R_{\textrm{SGE}}$ also reflects the spin diffusion
within the channel, which tends to shift the nonlocal signal maxima
towards higher gate voltages $|V_{g}|$ (see also Ref. \citep{Lin_2020}).
Most importantly, the very evidence of a maximum in $|R_{\textrm{SGE}}|$
as a function of carrier density shows that the spin-helical regime
is experimentally accessible. Moreover, the qualitative behavior of
the CSC efficiency (unlike the spin Hall angle \citep{Offidani_2018c})
is little sensitive to microscopic details of the impurity potential
\citep{Offidani_2017}. This provides extra confidence that the behavior
of the spin galvanic nonlocal signal in }Ref.\textcolor{black}{{} \citep{Benitez_2020}
is a fingerprint of emergent spin-helical Dirac fermions at the graphene/TMD
interface. On the other hand, a sign change in the SHE signal as the
back-gate voltage is swept at fixed carrier polarity as predicted
here (see Fig. \ref{fig:04}) is not evident in the experimental data
\citep{Benitez_2020}. The significant uncertainty in the nonlocal
signal at low temperatures renders a quantitative comparison between
theory and experiment more challenging. According to our theory, the
sign change occurs when $|\varepsilon|=|\lambda|\approx29$ meV. Incidentally,
the latter roughly coincides with the energy scale of electron-hole
puddles in the experiment $\delta E=\hbar v\sqrt{\pi\delta n}$ (with
$\delta n\approx2.5\times10^{11}\:\textrm{cm}^{-2}$ the residual
carrier density), which could smear out the features of the extrinsic
SHE predicted here. Therefore, the sign change in the SHA, if observed
in cleaner samples, could provide a direct measure of the BR SOC energy
scale. Both features, i.e. the maximum in the CSC efficiency and the
sign change in SHA, thus provide valuable transport fingerprints of
the spin-helical regime realized in graphene with proximity-induced
SOC. }

\textcolor{black}{We briefly comment on the validity of our microscopic
formulation. Close to the charge neutrality point, a full self-consistent
evaluation of the self-energy is required to account for the behavior
of the longitudinal charge conductivity. Because the weak-disorder
condition implies $\epsilon\tau_{\parallel}\gtrsim10$, which requires
not too low carrier density, one needs also a very large SOC to have
the regime I spanning a reasonable energy range. Notwithstanding,
the predicted sign change in the SHA at $|\varepsilon|\approx|\lambda|$
should be generally valid irrespective of the level at which the self-energy
is evaluated and may provide and experimental test of the theory.
A detailed self-consistent evaluation to extend our theory closer
to the charge neutrality point remains an important development for
future work, but is beyond the scope of the present paper.}

\section{Conclusions}

In conclusion, we developed a microscopic theory that \textcolor{black}{captures
the key features of coupled spin-charge transport in graphene-based
heterostructures, in which Dirac fermions experience proximity-induced
spin-valley coupling as well as Bychkov-Rashba effect due to interfacial
breaking of inversion symmetry. We have restricted ourselves to the
spin-helical regime realized at low electronic densities, characterized
by the locking of spin and pseudospin degrees of freedom. Such a simplification
has allowed us to derive, within the framework of the Eilenberger
equation for the quasiclassical Green's function, a single transport
equation capturing the low-energy behavior of dirty spin-orbit-coupled
Dirac fermions. The Eilenberger equation exhibits a scattering kernel,
which is derived within a T-matrix expansion by projecting the disorder
potential in the energy eigenstate at the Fermi energy. Both symmetric
and skew scattering features have been connected to the spin and pseudo
spin texture of the eigenstate through explicit analytical expressions.
The spin-charge transport response functions are described in terms
of longitudinal ($\tau_{\parallel}$) and transverse ($\tau_{\perp}$)
transport times. The transverse transport time is different from zero
only in the presence of both Bychkov-Rashba interaction and spin-valley
coupling, in accord with the exact covariant conservation laws of
the Dirac-Rashba model \citep{Milletari_2017}. Interestingly, the
SHE vanishes, then changing sign, when the Fermi energy crosses the
Bychkov-Rashba energy scale, irrespective of the value of the spin-valley
coupling. This is a consequence of the fact that the equilibrium spin
and pseudo spin texture, at this particular value of the Fermi energy,
implies a vanishing total spin angular momentum. In the spin-helical
energy range, the inverse spin galvanic effect has a maximum at the
edge of the spin-minority band. These features then represent the
fingerprints of the spin-helical regime and may provide a direct way
to estimate the proximity-induced SOC in graphene heterostructures
using spin precession measurements in nonlocal geometry that can disentangle
SHE and inverse spin galvanic signals \citep{Cavill_2020}. The comparison
with recent experimental results \citep{Benitez_2020} provides a
reasonable estimate for the Bychkov-Rashba SOC parameter. More importantly,
the energy range within which the theory is valid appears to be experimentally
accessible, so that the results presented in this paper can be put
to a test. }
\begin{acknowledgments}
R.R. and A.F. are grateful to Sergio O. Valenzuela for useful comments.
R.R. would like to thank Cosimo Gorini for useful discussions. A.F.
gratefully acknowledges the financial support from the Royal Society,
London through a Royal Society University Research Fellowship. 
\end{acknowledgments}

\appendix

\section{Details on gradient expansion}

To clarify the procedure, we can consider the simplest possible example:
free electrons in a perfect (no disorder) system and in the presence
of an electric field described by a scalar potential. We start from
the Dyson equation Eq.(\ref{eq:LR_Dyson}) and move to Wigner coordinates.
A convolution product $A(1,2)\otimes B(2,1')$ in Wigner space can
be written as

\[
(A\otimes B)(X,k)=e^{i(\partial_{X}^{A}\partial_{k}^{B}-\partial_{k}^{A}\partial_{X}^{B})}A(X,k)B(X,k)
\]

\noindent where the coordinates $(X,k)$ define the so-called mixed
representation. Here $k=(\mathbf{k,\omega})$. If $A(X,k)$ and $B(X,k)$
are slowly varying functions of $X$, the exponential can be expanded
order by order in the small parameter $\partial_{X}\partial_{k}\ll1$.
One can ultimately generates from the Eq.(\ref{eq:LR_Dyson}) an approximated
equation. In the lowest-order approximation of gradient expansion,
we have

\[
e^{i(\partial_{X}^{A}\partial_{k}^{B}-\partial_{k}^{A}\partial_{X}^{B})/2}\approx1+\frac{i}{2}(\partial_{X}^{A}\partial_{k}^{B}-i\partial_{k}^{A}\partial_{X}^{B}),
\]

\noindent so, beyond the standard semiclassical assumption that the
potential varies slowly in space and time on the scale set by $1/k_{F},1/\varepsilon$,
one has

\begin{eqnarray*}
-i\left[\hat{G}_{0}^{-1},^{\otimes}\hat{G}\right]\approx\qquad\qquad\qquad\\
\approx\partial_{T}\hat{G}-e\partial_{T}\Phi\partial_{\epsilon}G-\nabla_{\boldsymbol{p}}\hat{G}_{0}^{-1}\cdot\nabla_{\boldsymbol{R}}\hat{G}_{0}^{-1}\cdot\nabla_{\boldsymbol{k}}\hat{G}=\\
=\partial_{T}\hat{G}-e\partial_{T}\Phi\partial_{\epsilon}G+\boldsymbol{v}\cdot\nabla_{\boldsymbol{R}}\hat{G}+e\nabla_{\boldsymbol{R}}\Phi(\boldsymbol{R})\cdot\nabla_{\boldsymbol{k}}\hat{G}
\end{eqnarray*}

\noindent where $\hat{G}=\hat{G}(X,k)$, $\boldsymbol{v}=\boldsymbol{k}/m$,
$\Phi$ is the scalar potential as function of the Wigner variable
$\boldsymbol{R}$. From that, after some simple algebra, one can recover
the Eq.(\ref{eq:gradient_expansion}). \\

\section{On the detailed balance}

To see better how the detailed balance is obeyed in the picture presented
in this work, we have to start from $I_{in}$ and $I_{out}$ terms
(the Eq.(\ref{eq:I_in_expression}) and the Eq.(\ref{eq:I_out_expression})).
They read 
\begin{eqnarray*}
I_{in}=iu_{0}^{2}\int_{{\bf k'}}\Big(G_{\mathbf{k}}^{R}\overline{T}^{R}G_{{\bf k'}}^{K}\overline{T}^{A}-\overline{T}^{R}G_{{\bf k'}}^{K}\overline{T}^{A}G_{\mathbf{k}}^{A}\Big)
\end{eqnarray*}
and 
\begin{eqnarray*}
I_{out}=-iu_{0}^{2}\int_{{\bf k'}}\Big(G_{{\bf k'}}^{R}\overline{T}^{R}G_{\mathbf{k}}^{K}-G_{\mathbf{k}}^{K}G_{{\bf k'}}^{A}\overline{T}^{A}\Big).
\end{eqnarray*}

\noindent Now, we focus on $I_{out}$ term and project it in the basis
in which $G^{R(A)}$ (and then also $\overline{T}^{R(A)}$) is diagonal.
So, the $(ij)$-element reads

\begin{eqnarray*}
(I_{out})_{ij} & = & -iu_{0}^{2}\int_{{\bf k'}}\Big(G_{{\bf k'},i}^{R}\overline{T}_{i}^{R}G_{\mathbf{k},(ij)}^{K}-G_{\mathbf{k},(ij)}^{K}G_{{\bf k'},j}^{A}\overline{T}_{j}^{A}\Big)\\
 & = & -iu_{0}^{2}\int_{{\bf k'}}G_{\mathbf{k},(ij)}^{K}\Big(G_{{\bf k'},i}^{R}\overline{T}_{i}^{R}-G_{{\bf k'},j}^{A}\overline{T}_{j}^{A}\Big).
\end{eqnarray*}
One can rewrite it as 
\begin{eqnarray*}
(I_{out})_{ij} & = & -iu_{0}^{2}G_{\mathbf{k},(ij)}^{K}\Big(\frac{G_{{\bf k'},i}^{R}}{1-u_{0}G_{{\bf k'},i}^{R}}-\frac{G_{{\bf k'},j}^{A}}{1-u_{0}G_{{\bf k'},j}^{A}}\Big)\\
 & = & -iu_{0}^{2}G_{\mathbf{k},(ij)}^{K}\Big(G_{{\bf k'},i}^{R}-G_{{\bf k'},j}^{A}\Big)\overline{T}_{i}^{R}\overline{T}_{j}^{A}.
\end{eqnarray*}
From this result, we can recover the matrix relation for the $out-$term
being careful about the position of each element. In particular one
obtains

\begin{eqnarray*}
I_{out}=-iu_{0}^{2}\int_{{\bf k'}}\Big(G_{{\bf k'}}^{R}\overline{T}^{R}G_{\mathbf{k}}^{K}\overline{T}^{A}-\overline{T}^{R}G_{\mathbf{k}}^{K}\overline{T}^{A}G_{{\bf k'}}^{A}\Big).
\end{eqnarray*}
The latter expression has the same matrix structure of the $in-$relation,
as expected. 

\section{Effective potential in the $C_{6v}$-invariant model}

\label{details_W} We recall that in this case $\alpha=-\beta=-\varepsilon/\sqrt{\varepsilon(\varepsilon+2\lambda)}$,
$\gamma=-1$. Hence the normalization of the eigenvector (\ref{eq:generic_eigenvector})
reads 
\begin{equation}
1+\alpha^{2}+\beta^{2}+\gamma^{2}=4\frac{\varepsilon+\lambda}{\varepsilon+2\lambda}.
\end{equation}
Although the disorder potential does not have a dependence on the
scattering angle, its projection on the energy eigenstates gives rise
to such a dependence. To this end we define the scattering amplitude
for eigenstates at the Fermi surface with wave vector $\mathbf{k}=k_{F}\mathbf{n}$
and $\mathbf{k'}=k_{F}\mathbf{n'}$ where $\mathbf{n}=\left(\cos\theta,\sin\theta\right)$
and similarly for $\mathbf{n'}$
\begin{eqnarray}
p(\theta,\theta') & \equiv & \langle\Phi(k_{F},\theta)|\Phi(k_{F},\theta')\rangle.\nonumber \\
 & = & e^{i(\theta-\theta')}+\alpha^{2}+\beta^{2}+\gamma^{2}e^{-i(\theta-\theta')}\nonumber \\
 & = & \alpha^{2}+\beta^{2}+\mathbf{n}\cdot\mathbf{n'}\left(1+\gamma^{2}\right)\nonumber \\
 & + & i\mathbf{n}\times\mathbf{n'}\cdot\mathbf{z}\left(1-\gamma^{2}\right).\label{eq:scattering_amplitude_eigenstate}
\end{eqnarray}
One sees how the skew scattering in the last term in the right hand
side arises from the projection. In the $C_{6v}$-invariant model
when $\gamma=-1$ skew scattering is absent in lowest order in the
scattering amplitude and cannot appear in any successive order. To
evaluate the scattering kernel we then need the trace
\begin{eqnarray}
 &  & {\rm Tr}\left[P(\theta)P(\theta')\right]\nonumber \\
 & = & \frac{p\left(\theta,\theta'\right)p\left(\theta',\theta\right)}{\left(1+\alpha^{2}+\beta^{2}+\gamma^{2}\right)^{2}}\\
 & = & \frac{(\varepsilon+(\varepsilon+2\lambda)\cos(\vartheta))^{2}}{(\varepsilon-\lambda)^{2}},\nonumber 
\end{eqnarray}
where $\vartheta=\theta-\theta'$ as usual and in the last line we
used the expressions for $\alpha,\beta,\gamma$ given above. Finally,
by recalling that $N_{F}/N_{0}=(\varepsilon+\lambda)/\varepsilon$
and the definition (\ref{eq:total_Kernel_potential}) of the scattering
kernel, one obtains 
\begin{equation}
W(\vartheta)=\frac{1}{\tau}\frac{(\varepsilon+(\varepsilon+2\lambda)\cos(\vartheta))^{2}}{\varepsilon(\varepsilon+\lambda)},
\end{equation}
which is the result quoted in the main text.

\section{Expression of the effective potential and transport times in the
$C_{3v}$-invariant model}

\label{traces_expressions} We want to find the expressions of the
scattering kernel in the general case in terms of the eigenvector
coefficients defined in Eq.(\ref{eq:generic_eigenvector}). The Born
approximation and skew scattering contributions are defined in Eqs.
(\ref{eq:effective_kernel}) and (\ref{eq:skew_scattering_kernel}),
respectively. For the scattering kernel in the Born approximation
we need, apart from the eigenvector normalization factors, 
\begin{eqnarray*}
{\rm Tr}\left(P(\theta)P(\theta')\right) & = & p(\theta,\theta')p(\theta',\theta).
\end{eqnarray*}
By defining as usual $\vartheta=\theta-\theta'$ and taking the $(1-\cos(\vartheta))$-component
as requested by Eq.(\ref{eq:tau_par}) 
\begin{equation}
2\left(-\left(\gamma^{2}+1\right)\left(\alpha^{2}+\beta^{2}\right)+\left(\alpha^{2}+\beta^{2}\right)^{2}+\gamma^{4}+1\right).
\end{equation}
In a similar way for the scattering kernel relevant for the skew scattering
\begin{eqnarray*}
 &  & -i{\rm Tr}\left(\langle P(\theta'')\rangle\left[P(\theta),P(\theta')\right]\right)\\
 & = & 2{\rm Im}\int_{0}^{2\pi}\frac{d\theta''}{2\pi}p(\theta'',\theta)p(\theta,\theta')p(\theta',\theta'').
\end{eqnarray*}
After integrating over $\theta''$ and setting $\vartheta=\theta-\theta'$,
we may consider the $\sin(\vartheta)$ component as requested by Eq.(\ref{eq:tau_perp})
\begin{equation}
2\left(\gamma^{2}-1\right)\left(\alpha^{2}+\beta^{2}\right)\left(\alpha^{2}+\beta^{2}-\gamma^{2}-1\right).
\end{equation}
We finally have the transport times expressed in terms of the components
of the eigenvector, where we have inserted back the normalization
factors,
\begin{eqnarray*}
\frac{\tau}{\tau_{s}} & = & \frac{N_{F}}{N_{0}}\frac{2\left(\gamma^{4}+1+(\alpha^{2}+\beta^{2})(\alpha^{2}+\beta^{2}-\gamma^{2}-1)\right)}{(1+\alpha^{2}+\beta^{2}+\gamma^{2})^{2}},\\
\frac{\tau}{\tau_{a}} & = & g_{ss}\left(\frac{N_{F}}{N_{0}}\right)^{2}\frac{2\left(1-\gamma^{2}\right)\left(\alpha^{2}+\beta^{2}\right)\left(\alpha^{2}+\beta^{2}-\gamma^{2}-1\right)}{(1+\alpha^{2}+\beta^{2}+\gamma^{2})^{3}}.
\end{eqnarray*}

\section{Disorder parameters}

\label{disorder_estimates} In numerically evaluating the physical
observables we need to fix the strength of the disorder potential,
which enters in the expression of the scattering time and of the coupling
constant controlling the expansion about the Born approximation. They
are, after reinstating $\hbar$, 
\begin{eqnarray}
\frac{1}{\tau} & = & \frac{2\pi}{\hbar}N_{0}n_{imp}u_{0}^{2}=\frac{2\varepsilon}{\hbar}\left(\frac{u_{0}}{v\hbar}\right)^{2}n_{imp},\\
g_{ss} & = & 2\pi N_{0}u_{0}=2\frac{\varepsilon}{v\hbar}\frac{u_{0}}{v\hbar}.
\end{eqnarray}
The theory is valid in the limit $\varepsilon\tau/\hbar\gg1$ and
for $g_{ss}\ll1$. The quantity $u_{0}/(v\hbar)$ defines a typical
length associated with the scattering potential. For typical scattering
$u_{0}=1{\rm eV}{\rm nm}^{2}$, one has $u_{0}/(v\hbar)\sim10^{-9}{\rm cm}$.
For energies $\varepsilon\sim100{\rm meV}$, $\varepsilon/(v\hbar)\sim10^{6}{\rm cm}^{-1}$
and the weak scattering condition $g_{ss}\ll1$ is satisfied. With
impurity density $n_{imp}=10^{12}{\rm cm}^{-2}$, also the good metal
condition is verified. In the numerical plots we use the Rashba SOC
$\lambda$ as unit of energy. By indicating with $R=10^{-7}{\rm cm}$
the range of the impurity potential such that $u_{0}=UR^{2}$, we
may express the two above disorder parameters as 
\begin{eqnarray}
\frac{\hbar}{\lambda\tau} & = & 2\frac{\varepsilon}{\lambda}\left(\frac{U}{\lambda}\right)^{2}\left(\frac{\lambda R^{2}}{v\hbar}\right)^{2}n_{imp},\\
g_{ss} & = & 2\frac{\varepsilon}{\lambda}\frac{U}{\lambda}\frac{\lambda}{v\hbar}\frac{\lambda R^{2}}{v\hbar}.
\end{eqnarray}
We take $\lambda=10{\rm meV}$ and $v=10^{8}{\rm cm}\:\textrm{s}^{-1}$
so that 
\begin{eqnarray}
\frac{\hbar}{\lambda\tau} & = & 1.8\times10^{-3}\frac{\varepsilon}{\lambda}\\
g_{ss} & = & 1.8\times10^{-2}\frac{\varepsilon}{\lambda}.
\end{eqnarray}
The two above relations fix the value of the disorder parameters in
terms of the Fermi energy.

\end{document}